\documentclass[prd,nofootinbib,preprintnumbers,eqsecnum,superscriptaddress,showpacs]{revtex4-2}

\usepackage{graphicx}
\usepackage{amsmath}
\usepackage{amsfonts}
\usepackage{amsthm}
\usepackage{tikz}
\usepackage{subcaption}
\usepackage{bm}
\usepackage{soul}
\usepackage[sort&compress]{natbib}
\newtheorem*{theorem*}{Theorem}
\usepackage{float}
\usepackage{diagbox}
\usepackage{array}

\begin{document}

\title{Unraveling Generalized Parton Distributions Through Lorentz Symmetry \\and Partial DGLAP Knowledge}

\author{P.~Dall'Olio}
\affiliation{Dpto. Ciencias Integradas, Centro de Estudios Avanzados en Fis., Mat. y Comp., Fac. Ciencias Experimentales, Universidad de Huelva, Huelva 21071, Spain}
\author{F.~De Soto}
\affiliation{Dpto. Sistemas F\'isicos, Qu\'imicos y Naturales, 
Univ. Pablo de Olavide, 41013 Sevilla, Spain}
\author{C. Mezrag}
\affiliation{Irfu, CEA, Université Paris-Saclay, 91191, Gif-sur-Yvette, France}
\author{J.M.~Morgado Ch\'avez}
\affiliation{Irfu, CEA, Université Paris-Saclay, 91191, Gif-sur-Yvette, France}
\author{H.~Moutarde}
\affiliation{Irfu, CEA, Université Paris-Saclay, 91191, Gif-sur-Yvette, France}
\author{J.~Rodr\'{\i}guez-Quintero}
\affiliation{Dpto. Ciencias Integradas, Centro de Estudios Avanzados en Fis., Mat. y Comp., Fac. Ciencias Experimentales, Universidad de Huelva, Huelva 21071, Spain}
\affiliation{Irfu, CEA, Université Paris-Saclay, 91191, Gif-sur-Yvette, France}
\author{P.~Sznajder}
\affiliation{National Centre for Nuclear Research (NCBJ), 02-093 Warsaw, Poland}
\author{J.~Segovia}
\affiliation{Dpto. Sistemas F\'isicos, Qu\'imicos y Naturales, 
Univ. Pablo de Olavide, 41013 Sevilla, Spain}

\begin{abstract}
    Relying on the polynomiality property of generalized parton distributions, which roots on Lorentz covariance, we prove that it is enough to know them at vanishing- and low-skewness within the DGLAP region to obtain a unique extension to their entire support up to a D-term. We put this idea in practice using two methods: Reconstruction using artificial neural networks and finite-elements methods. We benchmark our results against standard models for generalized parton distributions. In agreement with the formal expectation, we obtain a very accurate reconstructions for a maximal value of the skewness as low as $20\%$ of the longitudinal momentum fraction. This result might be relevant for reconstruction of generalized parton distribution from experimental and lattice QCD data, where computations are for now, restricted in skewness.
\end{abstract}

\maketitle

\section{Introduction}

Introduced in the 1990s \cite{Mueller:1998fv,Ji:1996nm,Radyushkin:1997ki}, Generalized Parton Distribution (GPDs) are today at the core of hadron structure studies, both experimentally and theoretically.
This enthusiasm can be understood, as GPDs encode the multidimensional structure of hadrons \cite{Burkardt:2000za,Diehl:2002he} (2D in the transverse plane and 1D in longitudinal momentum space).
They are also connected to the energy-momentum tensor, allowing in principle to extract the contributions of quarks and gluons to the total angular momentum \cite{Ji:1996ek}, and also the contributions to the pressure and shear forces within the hadrons \cite{Polyakov:2002yz}.

Therefore, since the early 2000s, attempts to extract GPDs from experimental data have been performed (see the examples of \cite{Kumericki:2015lhb,Moutarde:2018kwr}). However the task remains hard for multiple reasons. 
First, GPDs are connected to experimental data of deep exclusive processes, like deep virtual Compton scattering (DVCS), which are much harder to measure than inclusive ones, connected to regular parton distribution functions (PDFs).
Further, the connection between experimental data and GPDs in standard deep exclusive processes like DVCS, reveals itself non invertible at fixed scale and mathematically ill-posed when evolution is turned on~\cite{Bertone:2021yyz,Moffat:2023svr}. 
The situation might be better for processes involving the production of an additional particle in the final state (see \cite{Boussarie:2016qop,Duplancic:2023kwe,Grocholski:2021man,Qiu:2022bpq,Qiu:2023mrm}).

The difficulty in extracting GPDs comes also from the fact that they have to obey a number of theoretical constraints whose fulfillment is \textit{a priori} not granted using generic modeling techniques. 
Among those, let us mention polynomiality \cite{Ji:1998pc,Radyushkin:1998bz} and positivity \cite{Radyushkin:1998es,Pire:1998nw,Diehl:2000xz,Pobylitsa:2002gw}.
A way to fulfill systematically both at the same time was introduced in \cite{Chavez:2021llq} based on the so-called covariant extension \cite{Chouika:2017dhe,Chouika:2017rzs}, and yielded the first evaluation of experimental feasibility of pion DVCS \cite{Chavez:2021koz}.
The covariant extension technique was originally developed to recover the Efremov-Radyushkin-Brodsky-Lepage (ERBL) kinematic region from the Dokshitzer-Gribov-Lipatov-Altarelli-Parisi (DGLAP) one, in a unique way, up to a $D$-term\footnote{The details about the ambiguities generated by the $D$-term can be found in ref. \cite{Chouika:2017dhe} and Sec.~\ref{sec:RTInv-th}.}. 
However, as we will show in this paper, its application is in fact more general and allows one to uniquely continue a GPD known solely in the vanishing and low-$\xi$ region to the entire kinematic domain. 
This application is important as it allows reconstruction for GPDs based on collider data (see for instance \cite{Dutrieux:2023qnz}) supplemented\footnote{The reader can refer to \cite{Riberdy:2023awf} for the impact of combining experimental and lattice data.} with lattice QCD data at vanishing $\xi$ (but finite values of $t$).
In such case, the $t$-dependent PDFs (or equivalently vanishing-$\xi$ GPDs) become key quantities, and they are typical matrix elements that can be assessed on the lattice (see for instance \cite{Karpie:2021pap,Bhattacharya:2022aob,Bhattacharya:2023nmv}).

In section 2, we start by recalling the mathematics behind the covariant extension, highlighting that both an incomplete $x$ and $\xi$ domain are allowed. In section 3, we check the actual feasibility using two different numerical techniques, Artificial Neural Networks (ANNs) and Finite-Elements Method (FEM), on two models. The first one is derived from lightfront wave functions computations \cite{Chouika:2017rzs} and the second one is the phenomenological parametrization by Goloskokov and Kroll \cite{Goloskokov:2005sd}. Finally in section 4, we conclude. 

\section{Polynomiality and inversion of the Radon transform}\label{sec:RTInv-th}

GPDs are defined as a lightfront projection of a non-diagonal hadronic matrix element of a bi-local operator \cite{Mueller:1998fv,Ji:1996nm,Radyushkin:1997ki}, and they are usually expressed in terms of three kinematic variables. Namely, the lightfront momentum fraction, $x$; the skewness or lightfront momentum fraction transfer, $\xi$; and the squared momentum transfer, $t$.  Concerning $x$ and $\xi$, the kinematic domain of GPD defines its support with $x,\xi \in [-1,1]$; the region $|x| \ge |\xi|$ ($|x| \leq |\xi|$) being denominated DGLAP (ERBL) where the GPD can be expanded in Fock space only involving states with 
the same number $N$ (different number $N$ and $N+2$) of particles\,\cite{Ji:1998pc}. 

This being given, it can be proven for any $m$-order GPD's Mellin moment that Lorentz symmetry implies \cite{Ji:1998pc,Radyushkin:1998bz,Polyakov:1999gs}  
\begin{align}\label{eq:poly}
\int_{-1}^{1} dx \, x^m H(x,\xi,t) = \sum_{k=0}^{m+1} c_k^{(m)}(t) \xi^k \;;    
\end{align}
where $H\left(x,\xi,t\right)$ denotes the unpolarized quark GPD, which we use here as illustration\footnote{In the case of polarized or transversity GPDs, the polynomiality property as expressed in Eq.~\eqref{eq:poly} must be modified accordingly \cite{Diehl:2003ny}.}. This property is widely dubbed \emph{polynomiality} and has both a deep origin and important implications. Particularly, if the GPD is known for all $x$ and $t$ in a given compact range of $\xi$, one can calculate Eq.\,\eqref{eq:poly}'s \textit{lhs} in there and, \emph{a priori}, determine the polynomial coefficients in \textit{rhs} for any $m$-order Mellin moment. It becomes thus clear that, rooting on Lorentz covariance and on the uniqueness of Mellin moments, the GPD appears defined in its entire support from its knowledge within a restricted kinematic domain. Moreover, the GPD reconstruction can be supplementary constrained by profiting the highly non-trivial interplay between the structure of the DGLAP and ERBL regions expressed by the polynomiality property in Eq.~\eqref{eq:poly}. In the following, we shall elaborate further on this, and exploit it, capitalizing specially on the related mathematical literature devoted to the incomplete data problem in \emph{computerized tomography}\,\cite{Natterer:2001}.         

To the purpose of this work, ultimately capitalizing on the result from Eq.\,\eqref{eq:poly}, the momentum transfer $t$ can be fixed and taken to be a constant parameter all throughout the analysis; any conclusion would then be irrespective of its particular value. Thus, for the sake of simplicity, it will be considered implicit from now on and omitted in the notation.   Furthermore, time reversal symmetry can be also invoked and seen to impose that any sensible GPD is an even function of $\xi$, thus entailing that no odd power is allowed in Eq.\,\eqref{eq:poly}: $c^{(m)}_{2n+1}=0$ for any integer $n$; and, particularly, $c^{(n)}_{n+1}=0$ for any even integer $n$. In the following we shall concentrate on the $\xi\geq 0$ region.

Then, as it has been carefully discussed in Ref.\,\cite{Chouika:2017dhe} (see also references therein), given a function $D(z)$ with support $z\in [-1,1]$, defined by its Mellin moments $\int_{-1}^1 dz \, z^m D(z) = c_{m+1}^{(m)}$, one can prove that any $m$-order Mellin moment of $H(x,\xi)-\mathrm{sgn}(\xi) D(x/\xi)$ is a polynomial of degree $m$ on $\xi$; otherwise said, it fulfills the Ludwig-Helgason consistency condition entailing that it is in the range of a Radon transform $\mathcal{R}$\,\cite{Hertle:1983}, 
\begin{align}\label{eq:RT0}
H(x,\xi)-\mathrm{sgn}(\xi) D\left(\frac{x}{\xi}\right) = \mathcal{R}F(x,\xi) = \int_{\Omega}d\beta d\alpha \, \delta\left(x-\beta-\alpha\xi\right) \, F\left(\beta,\alpha\right) \;; 
\end{align}
where $\Omega = \{ (\beta,\alpha) \in \mathbb{R}^2 / |\alpha|+|\beta| \leq 1 \}$ is the support for the distribution $F$  whose Radon transform $\mathcal{R}F$ is in the physical domain of $(x,\xi)$. $D$ is an odd function in its single argument, as it clearly comes from its definition in terms of Mellin moments 
which are zero for even order.   
Strictly speaking, the Ludwig-Helgason condition and the Radon transform are expressed\footnote{They admit a direct geometric interpretation representing the central case to which important mathematical literature has been thus far devoted\,\cite{Natterer:2001}.} in terms of the distance $s$ and the polar angle $\phi$  (\textit{e.g.}, see Ref.\,\cite{Chouika:2018mbk}), which can be mapped into the usual GPD variables $x$ and $\xi$ through the transformations $x=s/\cos{\phi}$ and $\xi=\tan{\phi}$, such that the $m$-order Mellin moment of $[H-\mathrm{sign}(\xi)D]/\cos{\phi}$ is an homogeneous polynomial of degree $m$ with terms $\cos^{m-k}{\phi} \sin^k{\phi}$. This change from mathematical to physical variables has no consequence for our purposes.  

One can then introduce a second distribution $G(\beta,\alpha)=\delta(\beta) D(\alpha)$ such that $\mathcal{R}G(\beta,\alpha)=D(x/\xi)/|\xi|$. Consequently, the GPD $H$ is given as $\mathcal{R}F\left(x,\xi\right)+\xi\mathcal{R}G\left(x,\xi\right)=H\left(x,\xi\right)$, with $F$ ($G$) being even (odd) in $\alpha$. Then, owing to its fulfilling of condition \eqref{eq:poly}, any sensible GPD can be represented as the Radon transform of the two 2-dimensional distributions $F$ and $G$, widely named double distributions (DDs). However, as exposed in Refs.\,\cite{Teryaev:2001qm,Tiburzi:2004qr}, the pair of DDs $(F, G)$ does not offer a unique representation for the GPD $H(x,\xi)$: there exists a family of transformations, usually called \emph{scheme} transformations, that make possible to redefine $F$ and $G$ for the same GPD. A judicious choice of the transformation (see for instance appendix B of Ref.\,\cite{Chouika:2017dhe}) allows for the two DDs to rely on one single DD $h$, the GPD then reading
\begin{align}\label{eq:RTP}
H(x,\xi) = (1-x) \mathcal{R}h(x,\xi) = (1-x) \int_{\Omega}d\beta d\alpha \, \delta\left(x-\beta-\alpha\xi\right) \, h\left(\beta,\alpha\right) \;,
\end{align}      
which corresponds to the so-called $P$-scheme\,\cite{Pobylitsa:2002vi}. At this point, it is worthwhile highlighting that, if $(1-x) \mathcal{R}h(x,\xi)$ fulfills the polynomiality condition \eqref{eq:poly}, 
\begin{align}\label{eq:RTP2}
\widetilde{H}(x,\xi) = (1-x) \mathcal{R}h(x,\xi)+ \mathrm{sign}(\xi) D^-(x/\xi)
\end{align}
also does, where $D^-(z)$ is any odd function in its single argument,  with support $z \in [-1,1]$. The latter is important for our purpose of facing the inverse problem, as $H$ and $\widetilde{H}$ only differ by the contribution from the line $\beta=0$ within the DD support, which impacts only on the GPD kinematic domain $|x| < |\xi|$. 

Let us, at this point, consider a GPD for which its knowledge is restricted to a given kinematic domain. Then, rooting on a proof of uniqueness for a DD whose Radon transform reproduces the GPD on such a restricted domain, the inversion of the Radon transform can be featured as a sensible procedure for a covariant extension from restricted to full GPD kinematic ranges. So has been suggested and illustrated in Refs.\,\cite{Chouika:2017dhe,Chouika:2017rzs,Chouika:2018mbk}, where GPDs known within their DGLAP kinematic domain were successfully extended to ERBL through the Radon transform of the DD previously obtained by inversion. 

More generally, the problem of inverting the Radon transform has been exhaustively studied in the mathematical context of computerized tomography\,\cite{Natterer:2001}; for which, in most realistic scenarios, the Radon transform is known only for hyperplanes belonging to a particular subset of the whole domain\footnote{In computerized tomography, this is the case when the particular geometry of the probing beam only allows to scan any 2-dimensional section of the problem object along a particular set of lines. And this is also the case in the context of the GPD covariant extension, where the original GPD data, expressed as the Radon transform of a DD, are only known for those lines belonging to the DGLAP region}. Although explicit inversion formulas remain mostly not available, several theorems prove the uniqueness of the inverse function, depending on the properties of the functions involved and given the data of the Radon transform corresponding to specific subsets of hyperplanes. This implies that if a particular function is found, whose Radon transform reproduces the data known for the subset of hyperplanes for which the theorems apply, then that function is guaranteed to be unique. The relevant theorem on which the GPD covariant extension relies  is the support theorem by Boman and Todd-Quinto\,\cite{boman:1987rad} that, for the sake of completeness, will be reproduced below:
\begin{theorem*}
Let us consider the Radon transform of a generalized function $f \in \mathcal{E'}(\mathbb{R}^n)$ along the hyperplane $z\cdot \theta =s$ ($z\in \mathbb{R}^n$, $\theta \in S^{n-1}$ and $s \in \mathbb{R}$)
\begin{equation}
\label{radongen}
\mathcal{R}f(\theta,s) = \int\limits_{z\cdot \theta=s}dz\, f(z).
\end{equation}
Let $W$ be an open unbounded connected subset of $S^{n-1}\times \mathbb{R}$, such that $\mathcal{R} f (\theta, s) = 0$ for $(\theta, s) \in W$. Then $f=0$ on $\bigcup \left\{z\cdot \theta=s | (\theta, s) \in W \right\}$.
\end{theorem*}
This theorem applies to $f \in  \mathcal{E'}(\mathbb{R}^n)$, \textit{i.e.} to distributions with compact support. It can therefore be applied to the Radon transform of a DD, whose domain is the compact skewed square $\Omega$. For this two dimensional case, the argument of the function corresponds to the DD variables, \textit{i.e.} $z \equiv (\beta, \alpha)$ and $\theta \in S^1 \equiv (\cos \phi, \sin \phi)$, which correspond to $(\theta, s)$, the Radon transform variables\footnote{Notice that although $s$ represents the distance between the hyperplane and the origin, the Radon transform is defined also for negative values of $s$, since it satisfies $\mathcal{R}f(\theta, s) = \mathcal{R}f(-\theta, -s)$} usually employed in mathematical literature and which convert to the standard GPD variables $(x,\xi)$ as made above explicit. In this way, the line $z\cdot \theta = s$ corresponds to $x-\beta -\alpha\xi=0$. 

What follows below aims at making transparent the physical implications of the above support theorem, exhibiting that it guarantees the uniqueness of the distribution $F(\beta,\alpha)$ from Eq.\,\eqref{eq:RT0}, if $\mathcal{R}F(x,\xi)$ is fixed by the knowledge of $H(x,\xi)$ on a subset of the DGLAP region. Notice first that Eq.\,\eqref{eq:RT0}'s \textit{lhs} also includes a term $D(x/\xi)$ which, anyhow, has to be evaluated outside its support for any DGLAP kinematic configuration and, hence, vanishes. However, this $D$-term raises a well-known ambiguity as, despite the uniqueness of $F(\beta,\alpha)$, the GPD $H(x,\xi)$ remains undetermined within the ERBL region by the impact of such a term which, as shown above, corresponds to a contribution with support only on the line $\beta=0$. The same is true for DDs in any scheme and, particularly, is made apparent by $P$-scheme Eqs.\,(\ref{eq:RTP},\ref{eq:RTP2}), which exhibit how DGLAP data cannot unambiguously fix the ERBL GPD. On the other hand, since $H(x, -\xi) = H(x, \xi)$, we can consider $\xi \geq 0$, implying $\phi \in [0,\frac{\pi}{2})$; and expose that the whole DGLAP region $|x| > \xi$, also expressed as $|s| > \sin \phi$, is not a connected set. However, this makes no limitation for the application of the theorem, because it can be separately applied for the two disjoint, open connected regions $\{|x| > \xi \} \cap \{x > 0 \}$ (particle contribution) and $\{|x| > \xi \} \cap \{x < 0 \}$ (anti-particle contribution).    

\begin{figure}[t]
\centering
\begin{minipage}[t]{0.4\textwidth}
    \begin{subfigure}[t]{0.5\textwidth}
        \centering
        \includegraphics[width=1.75\linewidth]{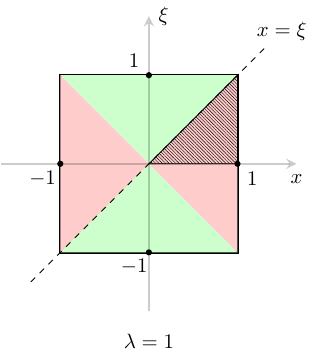}
    \end{subfigure}\\
    \begin{subfigure}[t]{0.5\textwidth}
        \centering
        \includegraphics[width=1.75\linewidth]{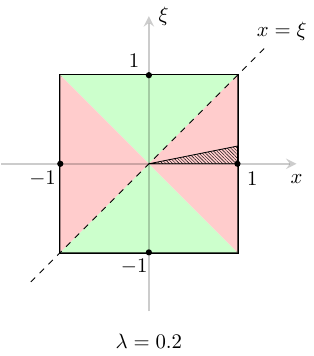}
    \end{subfigure}
\end{minipage}%
\begin{minipage}[c]{0.6\textwidth}
\includegraphics[scale=1.5]{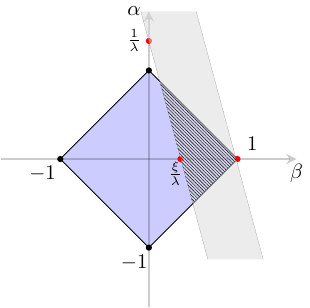}
\end{minipage}
\caption{\label{fig:regions} \small \textsc{Graphic illustration} . -- [\textsc{Left panel}] The square being a graphical representation of the domain of GPDs, green triangles correspond to the ERBL region and red areas to the DGLAP one. The hatched area represents, for different values of $\lambda$ in~\eqref{eq:lambda_region}, the portion of the overall GPD domain which is taken as input in the reconstruction procedure. [\textsc{Right panel}] The skewed square being the domain over which DDs have support, the Radon transform Eq.~\eqref{eq:RTP} can be interpreted as an integration over straight-lines, $\alpha=\left(x-\beta\right)/\xi$, with $\alpha_{0}=x/\xi$ and $\beta_{0}=x$ as alpha- and $\beta$-intercept, respectively. The support theorem guarantees the uniqueness for the DD within the region swept by all the straight-lines such that $(x,\xi) \in \mathcal{O}_\lambda$; namely, $\xi/\lambda < x \le 1$ with $0 \le \xi < \lambda$. Then, for any fixed $\xi$, one is left with parallel lines of slope $1/\xi$ bounded by $1/\lambda < \alpha_0 \le 1/\xi$ and, correspondingly, $\xi/\lambda < \beta_0 \leq 1$; schematically displayed in the plot by the shadowed area between the two parallels crossing the bounding intercepts (red circles). Therefore, when varying $\xi$ from $0$ to $\lambda$, the entire DD domain ($\beta > 0$) is covered.}
\end{figure}

Indeed, without lack of generalization, we can focus on the region $x>0$ and consider the following subsets (see the right plots of Fig.~\ref{fig:regions})
\begin{equation}\label{eq:lambda_region}
\mathcal{O}_\lambda = \{|x| > \xi \} \cap \{x > 0 \} \cap \{0 \le \xi <\lambda x \},
\end{equation}
where $\lambda\in\left(0,1\right]$. One should realize that $\{\beta>0\}=\bigcup\{z\cdot\theta=s|\left(\theta,s\right)\in\mathcal{O}_\lambda \}$. Namely, the set of lines on the $(\beta, \alpha)$ plane corresponding to $\mathcal{O}_\lambda$, \textit{i.e.} the lines $\alpha = (x-\beta)/\xi$, with $\xi \in (0, \lambda x)$, sweep the half plane $\beta > 0$ irrespective of the value of $\lambda$ (this is graphically illustrated by the right panel of Fig.\,\ref{fig:regions}). The support theorem therefore implies that if the GPD $H(x,\xi)$ is uniquely given on the DGLAP subset $\mathcal{O}_\lambda$, the DD is uniquely fixed on the half plane $\beta > 0$. By the same argument one can uniquely fix the DD on the $\beta < 0$ half plane by knowing the GPD on the corresponding $\mathcal{O}_\lambda$ subsets with $x<0$. 

Then, for any $\lambda \in (0,1]$, the DGLAP region $x > 0$ fixes uniquely the DD for all $\beta > 0$ and, separately, the DGLAP region $x < 0$ does the same for all $\beta < 0$. Furthermore, one can simply consider a positive $\alpha$ in both cases, owing to the even parity of the DDs $F$ (or $h$ in this scheme). On the other hand, as above discussed, the axis $\beta=0$ remains undetermined by the single implementation of DGLAP data. The choice of $\lambda=1$ entails that the GPD is uniquely determined, with the exception of the Radon transform of terms with support only on $\beta=0$, by its whole DGLAP knowledge. The covariant extension from DGLAP to ERBL kinematic ranges introduced and discussed in Refs.\,\cite{Chouika:2017dhe,Chouika:2017itz,Chouika:2017rzs,Chouika:2018mbk} relies on this remarkable output. However, 
more importantly and not thus far highlighted, is that the same is formally true for $0 < \lambda < 1$. Namely, the GPD covariant extension can be made from its knowledge on proper subsets of the DGLAP region where the skewness parameter is restricted to an arbitrarily small range, \textit{i.e.} $\xi\in\left[0,\lambda x\right]$, with a positive and arbitrarily small value of $\lambda$. Beyond its formal interest by itself, this is very relevant for physical purposes, since in actual experimental setups one usually has access only to small values of $\xi$; and can be also very helpful for the completion and GPD reconstruction from lattice data.   

In the following sections we will numerically examine the soundness of this conclusion by trying to invert the GPD Radon transform having as inputs its data on these restricted subsets $\mathcal{O}_\lambda$ and subsequently reconstructing the GPD on its whole domain.

\section{Proof of concept and feasibility}

If a kinematic completion of GPDs is formally possible, a proof of concept and feasibility yet remains to be studied. We therefore assess this problem by following two approaches: (1) Applying the FEM discussed and tested in Refs.\,\cite{Chouika:2017dhe, Chouika:2018mbk, Chavez:2021llq, MorgadoChavez:2022men}, Sec.~\ref{sec:FEM}; and (2) using ANNs for the extraction of DDs, Sec.~\ref{sec:ANNs}. The main purpose for this double implementation, with two very different methods to invert the Radon transform, stems from our aim of discarding systematic effects related to a particular inversion procedure. To this goal, we shall test the results obtained with both against two benchmarking models for GPDs\footnote{For further details about these two models see Appendix~\ref{app:Alg} and~\ref{app:GK}} (Tab.~\ref{tab:ThisWork}): The \textit{algebraic model} for the pion GPD first constructed in \cite{Chouika:2017rzs} and the renowned \textit{Goloskokov-Kroll model} \cite{Goloskokov:2005sd,Goloskokov:2006hr,Goloskokov:2007nt,Goloskokov:2009ia}.
A thorough inter-comparison of the two approaches, as they apply to our problem, is out of the scope of the present work.

\renewcommand{\arraystretch}{1.5}
\begin{table}[h]
    \centering
    \begin{tabular}{p{2.5cm}||p{2.5cm}|p{2.5cm}c}\hline\hline
       \diagbox[innerwidth=2.5cm,height=2\line]{Method}{Model}    & \centering Algebraic & \centering Goloskokov-Kroll & \\\hline
       FEM   & \centering Sec.~\ref{sec:FEMalg} & \centering Sec.~\ref{sec:FEMgk} & \\\hline
       ANN   & \centering Sec.~\ref{sec:ANNalg} & \centering Sec.~\ref{sec:ANNgk} & \\\hline\hline
    \end{tabular}
    \caption{Syllabus -- Summary of the models and strategies explored in this study.}
    \label{tab:ThisWork}
\end{table}
\renewcommand{\arraystretch}{1}

\subsection{Finite-Elements Method (FEM)}\label{sec:FEM}

A suitable numerical approach to the Radon transform inverse problem is to proceed through discretization and interpolation in the space of DDss, what in the following will be referred to as FEM-strategy. This methodology was first presented in \cite{Chouika:2017dhe} and further elaborated in \cite{Chavez:2021llq,MorgadoChavez:2022men} where its usefulness in the kinematic completion of GPDs was exposed, setting the ground for a pioneering exploratory study of DVCS on pions \cite{Chavez:2021llq}. As an outcome of these studies, a \texttt{C++} implementation of this tool is already available within the \texttt{PARTONS} framework \cite{Berthou:2015oaw}. However this work employs a slight extension of that version. We thus find it advantageous to briefly expose in the following the main ideas behind this program. The interested reader can find detailed discussions in \cite{Chouika:2017dhe, Chouika:2018mbk, Chavez:2021koz, MorgadoChavez:2022men}.

The starting point of the FEM-approach is to introduce a discretization of the DD domain, $\Omega =\bigcup_{e} \Omega_{e}$, and embed in it a set of interpolating polynomials such that the DD may be approximated within the discretized domain. The accuracy of this approximation is essentially determined by the choice of the interpolating functions as well as the mesh itself. Inspired by the method of finite-elements widespread in \textit{e.g.} the study of partial differential equations, we choose to perform piecewise polynomial interpolation\footnote{See App.~C in \cite{MorgadoChavez:2022men} and references therein for a detailed description of piecewise polynomial interpolation in two dimensions applied in this context.}, allowing to approximate the DD as
\begin{equation}\label{eq:Interpolation_e}
    h_{\textrm{FEM}}\left(\beta,\alpha\right)=\sum_{e}P_{e}\left(\beta,\alpha\right),
\end{equation}
where $P_{e}\left(\beta,\alpha\right)$ are the polynomial interpolants chosen for each element $\Omega_{e}$, which can in turn be expressed as 
\begin{equation}\label{eq:Interpolation_n}
    P_{e}\left(\beta,\alpha\right)=\left\lbrace\begin{array}{lcl}
    \displaystyle \sum_{k\in\Omega_{e}}v_{k}\left(\beta,\alpha\right)h_{k} &,&\textrm{for }\left(\beta,\alpha\right)\in\Omega_{e},   \\
    \\
    \displaystyle 0 &,&\textrm{otherwise},
    \end{array}\right.
\end{equation}
where $k$ labels the items of a set of interpolation nodes $n_{j}\equiv\left(\beta_{j},\alpha_{j}\right)$ appropriately distributed over $\Omega_{e}$; and where $v_{j}\left(\beta,\alpha\right)$ represents the Lagrange interpolating polynomials associated to the node $n_{j}$. In this work we take them to be degree-two polynomials\footnote{This feature constitutes the main upgrade of our implementation with respect to that of Refs.~\cite{Chouika:2017dhe,Chouika:2018mbk,Chavez:2021llq,MorgadoChavez:2022men} where degree zero and one interpolating polynomials where employed.}. Finally, $h_{j}$ is the value of the DD at the interpolation node $n_{j}$, so we may write the integral problem of Eq.~\eqref{eq:RTP} as
\begin{equation}
    H\left(x,\xi\right)=\sum_{j}\left[\int_{\Omega_{e}}d\beta d\alpha\delta\left(x-\beta-\alpha\xi\right)v_{j}\left(\beta,\alpha\right)\right]h_{j}=\sum_{j}\mathcal{R}v_{j}\left(x,\xi\right)h_{j}.
\end{equation}

Importantly, because the interpolating basis functions $v_{j}$ are simple second order polynomials, the integrals can be evaluated in closed form, turning the original problem into a set of algebraic equations connecting the DD and its associated GPD. Furthermore, choosing pairs $\left(x_{i},\xi_{i}\right)$, a system of algebraic equations can be built to represent the Radon transform problem:
\begin{equation}\label{eq:RT_matrixsys}
    \left(
    \begin{array}{c}
    \displaystyle H\left(x_{1},\xi_{1}\right)  \\
    \displaystyle \vdots \\
    \displaystyle H\left(x_{N},\xi_{N}\right)  \\
    \end{array}\right)=\left(
    \begin{array}{ccc}
    \displaystyle \mathcal{R}v_{1}\left(x_{1},\xi_{1}\right) & \displaystyle \cdots & \displaystyle \mathcal{R}v_{n}\left(x_{1},\xi_{1}\right) \\
    \displaystyle \vdots                                    & \displaystyle \ddots & \displaystyle \vdots \\
    \displaystyle \mathcal{R}v_{1}\left(x_{N},\xi_{N}\right) & \displaystyle \cdots & \displaystyle \mathcal{R}v_{n}\left(x_{N},\xi_{N}\right)
    \end{array}\right)
    \left(\begin{array}{c}
    \displaystyle h_{1}  \\
    \displaystyle \vdots \\
    \displaystyle h_{n}  
    \end{array}\right).
\end{equation}

In particular, one may choose the sampling points $\left(x_{i},\xi_{i}\right)$ within the DGLAP region, where the GPD is assumed to be known. Or even better, on a subset of that domain such as the one considered in Eq.~\eqref{eq:lambda_region}. When doing so, a system of algebraic equations where the only unknowns are the values of the DD at the interpolation nodes arises. As explained in \cite{Chavez:2021llq} its solution can be found using a least squares minimization
\begin{equation}\label{eq:RT_sol}
    \bm{h}_{\textrm{FEM}}=\left(\bm{\mathcal{R}}^{T}\bm{\mathcal{R}}\right)^{-1}\bm{\mathcal{R}}^{T}\left.\bm{H}\right|_{\textrm{DGLAP}},
\end{equation}
where boldface letters represent the vectors (matrices) of Eq.~\eqref{eq:RT_matrixsys} in compact form. Thereupon, one may finally reconstruct the DD according to Eq.~\eqref{eq:Interpolation_e} (or Eq.~\eqref{eq:Interpolation_n}) and employ it to evaluate the GPD within the ERBL region through a Radon transform (in its discretized version):
\begin{equation}\label{eq:FEM-covext}
    \left.\bm{H}\right|_{\textrm{ERBL}}=\widetilde{\bm{\mathcal{R}}}\bm{h}_{\textrm{FEM}}=\widetilde{\bm{\mathcal{R}}}\left(\bm{\mathcal{R}}^{T}\bm{\mathcal{R}}\right)^{-1}\bm{\mathcal{R}}^{T}\left.\bm{H}\right|_{\textrm{DGLAP}}
\end{equation}
where $\widetilde{\bm{\mathcal{R}}}$ is a matrix representation of the Radon transform integral operator built following the procedure described before and evaluated for $\left(x_{i},\xi_{i}\right)$ chosen within the ERBL region.

\begin{figure}[t]
\begin{subfigure}{.5\textwidth}
  \centering
  \includegraphics[width=0.95\linewidth]{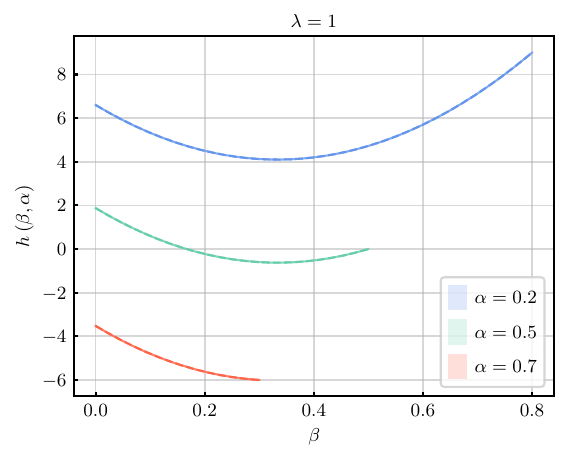}
\end{subfigure}%
\begin{subfigure}{.5\textwidth}
  \centering
  \includegraphics[width=0.95\linewidth]{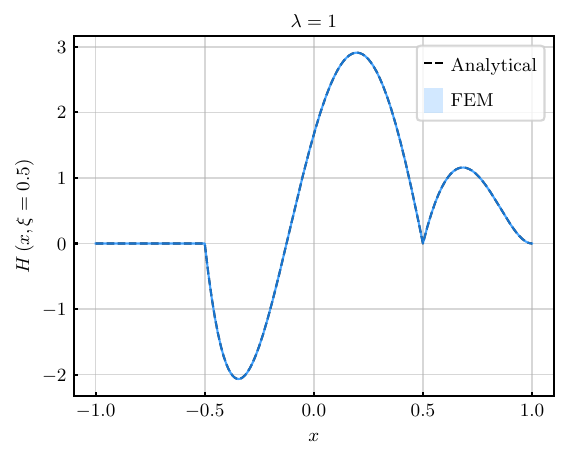}
\end{subfigure}
\caption{\small \textsc{(FEM) Algebraic model} -- The left panel shows the DD $h_{\textrm{FEM}}(\beta, \alpha)$ obtained through the procedure described in Sec.~\ref{sec:FEM} (solid lines), displayed as a function of the $\beta$ variable for three different values of $\alpha=\{0.2, 0.5, 0.7\}$ ($0\leq \beta \leq 1-\alpha$). For comparison, the exact analytical result of Eq.~\eqref{eq:alg_dd} is drawn through dashed lines. The sampling of the DGLAP region corresponds to $\xi\in\left[0,\lambda x\right]$ with $\lambda=1$ (the entire domain). The right plot displays the GPD $H\left(x,\xi\right)$ at the illustration value $\xi=0.5$, obtained as the Radon transform of the DD shown in the left, compared to the analytical expression, Eqs.~\eqref{eq:alg_dglap}~and~\eqref{eq:alg_erbl}.}
\label{fig:alg_fem_l1}
\end{figure}

In this way, a FEM-inspired strategy allows to find a solution for the inverse Radon transform problem. This approach being grounded on the theoretical construction of Sec.~\ref{sec:RTInv-th}, the covariant extension of GPDss from the DGLAP to the ERBL region turns feasible. Nonetheless, even if the input DGLAP GPD is known to arbitrary precision\footnote{Of course that situation is unlikely to be encountered in a realistic scenario. As an illustration one can consider the case of study where the input DGLAP GPD is extracted from simulations on the Lattice, thus each point being affected by the corresponding errors. In that case one must properly account for error propagation from the DGLAP to the ERBL region as driven by Eq.~\eqref{eq:FEM-covext}. However, the extraction of GPDs from the Lattice as well as their experimental assessment are still at an early stage; and more importantly the sources of this kind of errors have little to do with our approach. On the other hand, our numerical solution of an ill-posed problem such as the inverse Radon transform itself introduces systematic errors for which we may account without regard to the nature of the feeding distribution. They are precisely the sources and magnitudes of these last uncertainties which we assess in this work.}, the procedure is affected by uncertainties. Provided that the solution to our problem exists (Sec.~\ref{sec:RTInv-th}), the construction of this section reveals ($1$) the choice of interpolating polynomials, ($2$) the mesh definition (in particular its size) and ($3$) the number/distribution of sampling lines $\left(x_{i},\xi_{i}\right)$ in Eq.~\eqref{eq:RT_matrixsys} as the main sources of possible deviation from the actual solution. In practice, we have tested three types of polynomial interpolants: Zero, first and second order polynomials. Optimization in the sense of trading performance of the implementation for accuracy of the solution lead us to use degree-two polynomials for all calculations. With regard to the mesh, a \textit{Delaunay} triangulation was built over the DD domain \cite{Chouika:2017dhe,Chavez:2021llq}, the average size of the elements being fixed to a model-specific optimal value found as to balance precision (expected for finer meshes) and stability (precluded in the same direction) \cite{Naterer:1977fem}. Coincidentally, the meshes revealing optimal for the study of the algebraic- and GK-model where found to be identical and made up from seventeen elements of average area $0.03$ (arbitrary units). Finally, the number $N$ of sampling lines was set according to the criterion of \cite{Chavez:2021llq} where it was argued that the actual answer reveals in the least squares solution Eq.~\eqref{eq:RT_sol} when $N_{sample}$ is taken to be at least four times the number of elements in the mesh. Here we took $N_{sample}=6n$. These three sets of parameters being thus fixed, the solution to the inverse Radon transform problem can be found. From that point on there remains one single parameter whose effect on the solution has to be assessed: The distribution of sampling lines within the (restricted) DGLAP region. Here we follow once again the prescription of \cite{Chavez:2021llq} and decide to draw them randomly following a uniform distribution $x,\xi/x\in\left[0,1\right]$. Therefore, the distribution of lines leading to the solution of the inverse Radon transform varies among runs. In lack of a criterion to fix them, such constitutes the main source of statistical uncertainty in our calculation. Here we decide to estimate it through the production of replicas following an strategy similar to that in \cite{Dutrieux:2021wll}: We generate $250$ sets of sampling lines and solve our problem for each of them. Thus one is given with the corresponding number of solutions to our problem, say DDss, called \textit{replicas}. Then, for a given point, say the interpolation nodes, we check for outliers following a $5\sigma$-rule \cite{Ilyas:2019threesigma}. Finally, the replicas identified as outliers in any of the points are removed from the uncertainty estimate, which is given by the standard deviation of the population of replicas. Accordingly, all our results (Figs.~\ref{fig:alg_fem_l1}~to~\ref{fig:gk_fem_l05}) are given as a solid line, representing the mean value; and the corresponding one-sigma band.

\subsubsection{Algebraic model}\label{sec:FEMalg}

\begin{figure}[htb]
\begin{subfigure}{.5\textwidth}
\begin{subfigure}{\textwidth}
  \centering
  \includegraphics[width=0.95\linewidth]{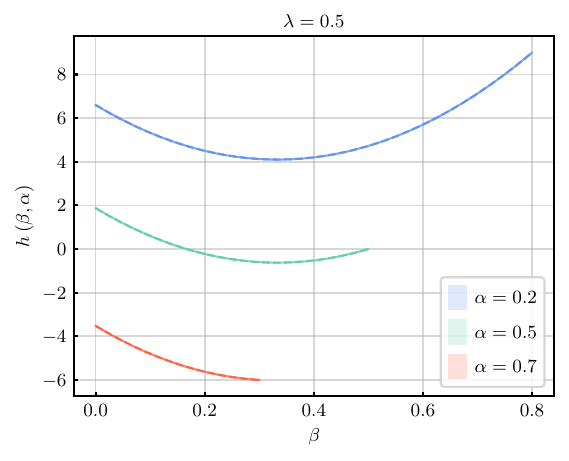}
\end{subfigure}
\begin{subfigure}{\textwidth}
  \centering
  \includegraphics[width=0.95\linewidth]{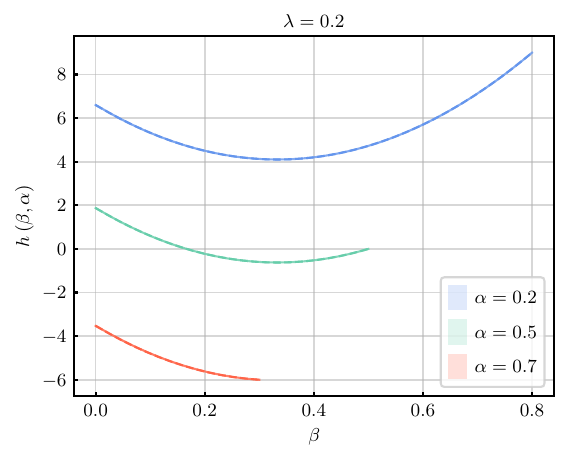}
\end{subfigure}
\end{subfigure}%
\begin{subfigure}{.5\textwidth}
\begin{subfigure}{\textwidth}
  \centering
  \includegraphics[width=0.95\linewidth]{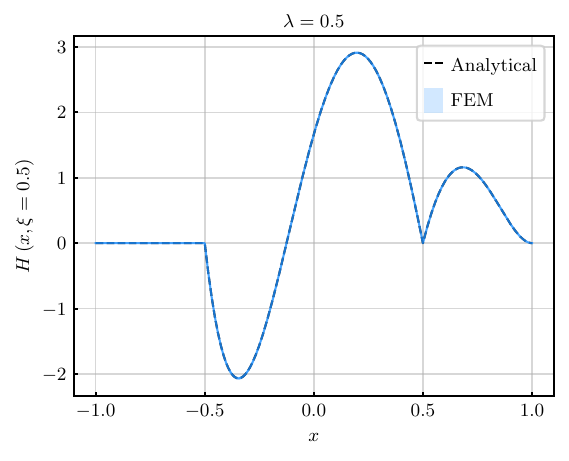}
\end{subfigure}
\begin{subfigure}{\textwidth}
  \centering
  \includegraphics[width=0.95\linewidth]{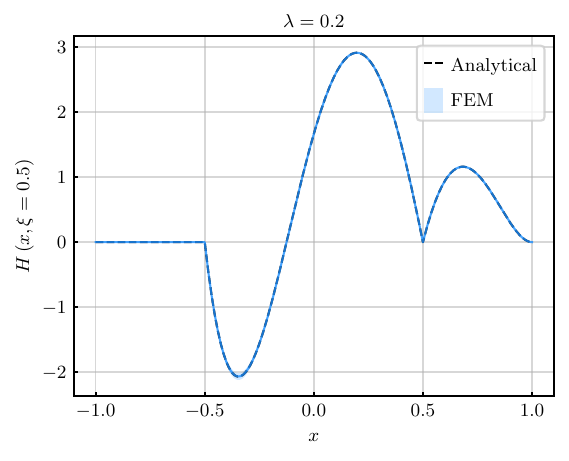}
\end{subfigure}
\end{subfigure}%
\caption{\small \textsc{(FEM) Algebraic model} -- The same of Fig.\,\ref{fig:alg_fem_l1} with a DGLAP sampling given by $\xi\in\left[0,\lambda x\right]$ and $\lambda = 0.5$ (top) and $0.2$ (bottom).}
\label{fig:alg_fem_l05}
\end{figure}

Employing the setup described in the previous paragraphs, the algebraic model can be extended from the DGLAP region, cf. Eq.~\ref{eq:alg_dglap}, to the ERBL domain. As in Sec.~\ref{sec:ANNs}, the case $\xi=0.5$ has been used to illustrate our findings. Figs.~\ref{fig:alg_fem_l1} and \ref{fig:alg_fem_l05} shows the results obtained as the solution to our problem using the region $\mathcal{O}_{\lambda}$ as the input domain. Three values of $\lambda$ have been explored, ranging for $\lambda=1$ (corresponding to the entire DGLAP region) down to $\lambda=0.5$ and $0.2$. In all cases, the agreement between the actual GPD and the reconstructed form is astonishing, showing that indeed the FEM strategy allows to perform the covariant extension of GPDs from the DGLAP to the ERBL region with a minimal knowledge of the input GPD. These results agree with those found in Sec.~\ref{sec:ANNs} using the ANN implementation, again supporting the thesis of this study.

The accuracy of the solution is remarkably good. We can get a better grasp about it by looking at the reconstructed DD $h_{\textrm{FEM}}\left(\beta,\alpha\right)$ against the expectation, Eq.~\eqref{eq:alg_dd}: left panels of Figs.~\ref{fig:alg_fem_l1} and \ref{fig:alg_fem_l05}. Again, both are essentially indistinguishable. The reason for that resides in the structure of the DD generating the algebraic model, which is a second order polynomial in the kinematic variables, Eq.~\eqref{eq:alg_dd}; just as they are the interpolants $P_{e}\left(\beta,\alpha\right)$ introduced within the FEM strategy. In that way, the exact solution is accidentally put by construction in the range of the discrete Radon transform of Eq.~\eqref{eq:RT_matrixsys}, meaning that in the absence of noise, the exact result can be found. In modern terms, this situation resembles that found in Bayesian reconstruction when a default model is introduced as prior information. In the language of Bayes theorem, such a default model represents the most probable answer in the absence of any data. When the chosen default model indeed represents the input data, the functional space is strongly dumped, and the chances of finding an accurate solution grow.

\subsubsection{Goloskokov-Kroll model}\label{sec:FEMgk}

\begin{figure}[htb]
\begin{subfigure}{.5\textwidth}
  \centering
  \includegraphics[width=0.95\linewidth]{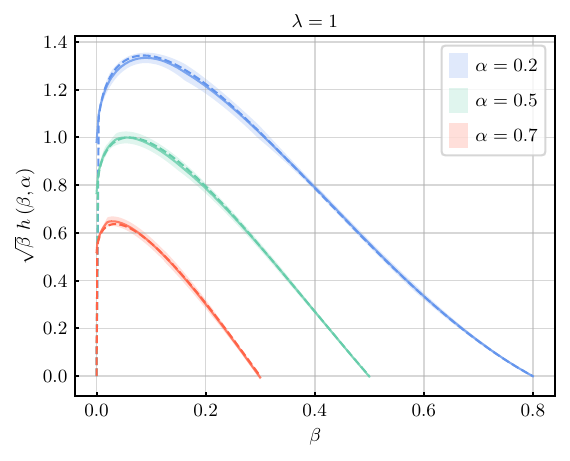}
\end{subfigure}%
\begin{subfigure}{.5\textwidth}
  \centering
  \includegraphics[width=0.95\linewidth]{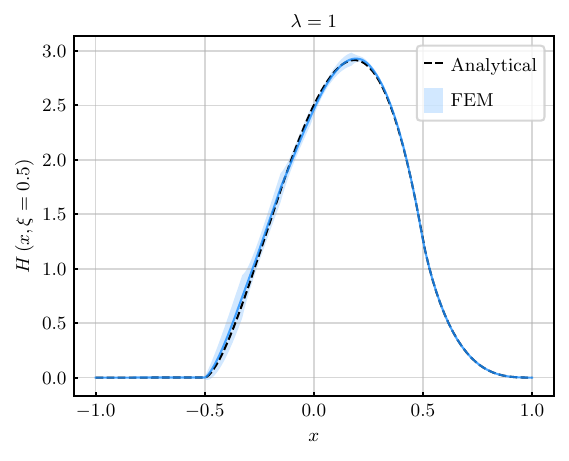}
\end{subfigure}
\caption{\small \textsc{(FEM) GK model} -- The left panel shows the DD $h_{\textrm{FEM}}(\beta, \alpha)$ obtained through the procedure described in Sec.~\ref{sec:FEM} (solid lines), displayed as a function of the $\beta$ variable for three different values of $\alpha=\{0.2, 0.5, 0.7\}$ ($0\leq \beta \leq 1-\alpha$). For comparison, the exact analytical result of Eq.~\eqref{eq:dd_gk} is drawn through dashed lines. The sampling of the DGLAP region corresponds to $\xi\in\left[0,\lambda x\right]$ with $\lambda=1$ (the entire domain). The right plot displays the GPD $H\left(x,\xi\right)$ at the illustration value $\xi=0.5$, obtained as the Radon transform of the DD shown in the left, compared to the analytical expression, Eqs.~\eqref{eq:gk_dglap},\eqref{eq:gk_erbl}.}
\label{fig:gk_fem_l1}
\end{figure}

\begin{figure}[htb]
\begin{subfigure}{.5\textwidth}
\begin{subfigure}{\textwidth}
  \centering
  \includegraphics[width=0.95\linewidth]{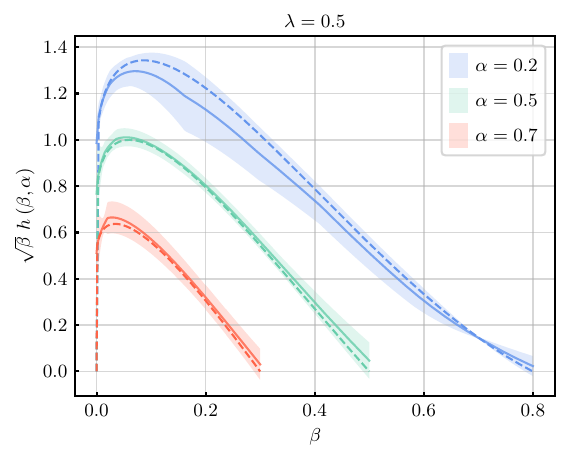}
\end{subfigure}
\begin{subfigure}{\textwidth}
  \centering
  \includegraphics[width=0.95\linewidth]{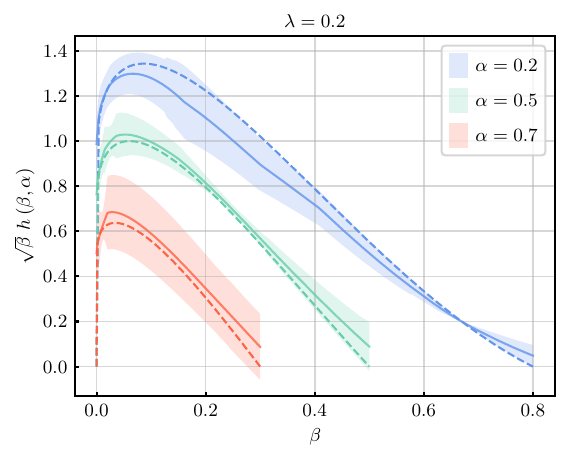}
\end{subfigure}
\end{subfigure}%
\begin{subfigure}{.5\textwidth}
\begin{subfigure}{\textwidth}
  \centering
  \includegraphics[width=0.95\linewidth]{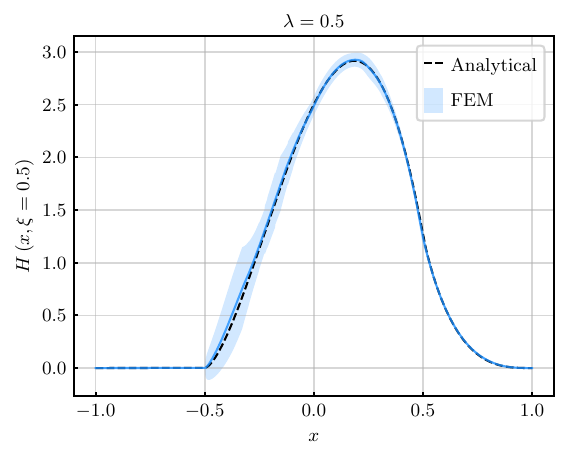}
\end{subfigure}
\begin{subfigure}{\textwidth}
  \centering
  \includegraphics[width=0.95\linewidth]{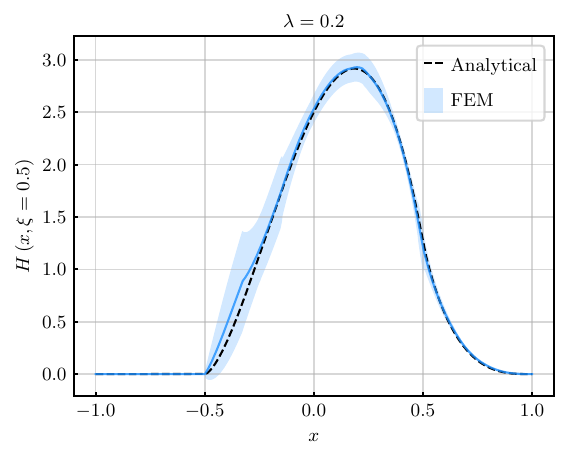}
\end{subfigure}
\end{subfigure}%
\caption{\small \textsc{(FEM) GK model} -- The same of Fig.\,\ref{fig:gk_fem_l1} with a DGLAP sampling given by $\xi\in\left[0,\lambda x\right]$, with $\lambda = 0.5$ (top) and $0.2$ (bottom).}
\label{fig:gk_fem_l05}
\end{figure}

With a proof of concept for the usefulness of the FEM construction in solving the inverse Radon transform problem in a particularly favorable case, challenging our implementation in more complicated scenarios becomes crucial. The case of the GK model constitutes an ideal test ground for three reasons: First, as remarked in Sec.~\ref{app:GK}, the GK model is widely employed in phenomenological analyses of GPDs. Second, the coincidence between the basis functions and the actual expression observed in the algebraic model is no longer found. And third, the DD giving raise to the GK model exhibits an integrable divergence at low values of $\beta$, challenging even more our implementation. It is precisely this later feature that has triggered a slight modification of the construction employed in the case of the algebraic model, redefining the approximating DD, $h_{\textrm{FEM}}$ as
\begin{equation}\label{eq:dd_fem_precond}
    h_{\textrm{FEM}}\left(\beta,\alpha\right)=r\left(\beta\right)\sum_{e}P_{e}\left(\beta,\alpha\right),\qquad r\left(\beta\right)\equiv \frac{1}{\sqrt{\beta}},
\end{equation}
so as to smooth the divergent behavior of $h_{\textrm{GK}}$ when applying the FEM reconstruction.

In practice we have found this redefinition to be key in stabilizing the problem's solution. Because the behavior of the underlying DD is generally unknown, this observation might seem a curse of the present approach. However, we believe that to be only apparent. Two main arguments can be highlighted here: On the one hand, although DDs are generally unknown, one can always rely on the behavior of the input DGLAP GPD to infer information about that of the structure of the underlying DDs.
As a simple illustration, one expect that the PDFs present a diverging behavior in the the limit $x\to 0$.
One may argue, for instance using the Goloskokov-Kroll model, that such low-$x$ divergences translate into a similar behavior of the DD as $\beta$ approaches zero. Moreover, these singularities are expected to be integrable in the case of valence distributions (as those being considered here, thus triggering our choice $r\left(\beta\right)=1/\sqrt{\beta}$). A second argument is of a practical nature. Indeed, the relevance of the smoothing suggested in Eq.~\eqref{eq:dd_fem_precond} can be traced back to the structure of the interpolating basis functions: Here they are simple degree-two polynomials which are unlikely to accurately piecewise-approximate rapidly growing functions such as $h_{\textrm{GK}}$ in the limit of small $\beta$. In contrast, the approximating functions employed in our ANN approach (where no special attention was payed to this feature) are more complicated functions of the kinematic variables $\left(\beta,\alpha\right)$, thus hinting a modification of the interpolants $P_{e}$ to be a general possible solution to this problem. Of course, there is no reason why not to upgrade the interpolating polynomials $P_{e}$ to more sophisticated structures capable of accounting for these kind of effects; however we prefer here to keep things simple and proceed using Eq.~\eqref{eq:dd_fem_precond}.

This being clarified, we can operate as before and explore the covariant extension of the GK model from the DGLAP to the ERBL region. Again we use $\xi=0.5$ as an illustration for our results; we study three possible sampling regions $\mathcal{O}_{\lambda}$, with $\lambda=1,0.5,0.2$, as well. The corresponding results are shown in the right panels of Figs.~\ref{fig:gk_fem_l1} and \ref{fig:gk_fem_l05}. Again, the overall fidelity of the reconstructed GPD is noticeable. However, this time we see a growing deterioration, not that substantial at the level of the mean value but at that of the uncertainties, as the area covered by the input data set decreases. In particular, deviations seem to be more apparent in the region $-\xi<x<0$. This can be traced back to numerical instabilities triggered by the smoothing factor $\sqrt{\beta}$ introduced in Eq.~\eqref{eq:dd_fem_precond}. Indeed, integration over $\beta$ translates (by means of the delta distribution in Eq.~\eqref{eq:RT0}) into $\sqrt{x-\alpha\xi}$, which shall be bounded from below. However, for values $x<0$, if numerical precision languishes, the argument might show very small while non-zero values, that being at the origin of the observed impaired behavior.

Further insights can be obtained by again looking at the ``intermediate'' DD (left panels of Figs.~\ref{fig:gk_fem_l1} and \ref{fig:gk_fem_l05}). The results show general good agreement, specially when the entire DGLAP region is given as input information. However, the reconstructed signal rapidly deteriorates as $\lambda$ decreases. Nevertheless, those deviations wash out after integration to reach the GPD domain, similarly to what will be seen in Sec.~\ref{sec:ANNs}. This is indeed a very noticeable feature, as the quantity relevant for phenomenology and the one in Lattice QCD is, indeed, the integrated DD; \textit{i.e.} the GPD.

\subsection{Artificial Neural Newtworks (ANNs)}\label{sec:ANNs}

An entirely different approach to tackle the inverse Radon transform problem consists in fitting the DD by means of an ANN with one hidden layer. The advantages of this strategy are twofold: Firstly, the universal approximation theorem \cite{HornikEtAl89} guarantees that this simple network architecture is able to approximate, at arbitrary precision for sufficiently large width, any compactly supported continuous function, therefore providing an unbiased parametrization of the DD; secondly, the optimization algorithms that are used to train the network come equipped with regularization methods, introduced to avoid overfitting, that are particularly useful in this context to overcome the ill-posedness of the Radon transform inverse problem\,\cite{Bishop1995, WagerSidaLiang2013}.

\begin{figure}
\centering
\includegraphics[width=0.33\linewidth]{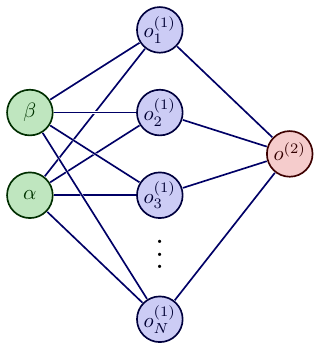}
\caption{\small ANN architecture with one hidden layer that parametrizes the DD $h(\beta, \alpha)$. Here $o^{\left(L\right)}_{i}$ stem for output of the $i$-th neuron in the $L$-th layer of the network. The zeroth layer is implicitly identified as the input one.}
\label{fig:ann}
\end{figure}

The DD $h(\beta, \alpha)$ is therefore approximated by the final output of the ANN whose architecture is sketched in Fig.~\ref{fig:ann}, with the two variables $\beta$ and $\alpha$ as input \textit{features}. As a function of the input variables, the network's output is denoted $h_{\textrm{ANN}}(\beta, \alpha)$. The explicit functional form, in terms of the weights ($w$) and biases ($b$) parameters is:
\begin{align}
\label{eq:ANNform}
h_{\textrm{ANN}}(\beta, \alpha) \equiv o^{(2)} &= \sum_{i=1}^N w^{(2)}_i o^{(1)}_i + b^{(2)} \\
&= \sum_{i=1}^N w^{(2)}_i\left[\sigma \left(w^{(1)}_{\beta i} \beta + w^{(1)}_{\alpha i} \alpha' + b^{(1)}_i \right) +  \sigma \left(w^{(1)}_{\beta i} \beta - w^{(1)}_{\alpha i} \alpha' + b^{(1)}_i \right)\right]+ b^{(2)}, \nonumber
\end{align}
where a re-scaled variable $\alpha'=\alpha/1-|\beta|$ has been introduced, that takes values between -1 and 1 (since $|\alpha|+|\beta|\leq 1$) which is supposed to better perform as an input to the sigmoid activation function $\sigma(x) = (1+e^{-x})^{-1}$. $N$ is the number of hidden neurons which is tuned in order to obtain the best possible approximation. The total number of learning parameters is $4N+1$, divided between $3N$ weights and $N+1$ biases. Notice that in Eq.~\eqref{eq:ANNform} the traditional hidden output has been augmented with an identical term with $w_\alpha \rightarrow - w_\alpha$, as it was already implemented in \cite{Dutrieux:2021wll}, which explicitly enforces the parity condition $h_{\textrm{ANN}}(\beta, -\alpha) = h_{\textrm{ANN}}(\beta, \alpha)$.

The network is trained using GPD data. Actually, the inputs for the algorithm are the GPD variables $\left(x_{i}, \xi_{i}\right)$, randomly chosen with a uniform probability distribution from the DGLAP region. Each pair of values $\left(x_{i}, \xi_{i}\right)$ corresponds to a line $\alpha = -\beta/\xi_{i} + x_{i}/\xi_{i}$ in the $\beta$--$\alpha$ plane, along which the Radon transform has to be evaluated (integrating $h_{\textrm{ANN}}(\beta, \alpha)$) to produce a predicted GPD value $\widehat{H}\left(x_{i}, \xi_{i}\right)$ that is to be compared with the true one $H\left(x_{i}, \xi_{i}\right)$. The network depicted in Fig.~\ref{fig:ann} may therefore be considered as embedded in a larger one, with two extra layers: An input layer that for each $(x_i, \xi_i)$ outputs the batch of $(\beta_k, \alpha_k)$ values positioned along the corresponding line\footnote{The size of the batch being the number of nodes whose values depend on the quadrature routine used to perform the numerical integration.}; and a final output layer that combines the batch of the $h_{\textrm{ANN}}\left(\beta_{k}, \alpha_{k}\right)$ values to produce $\widehat{H}\left(x_{i},\xi_{i}\right)$.

The network parameters are updated at each iteration, using the adaptive gradient descent \textit{Adam} algorithm\footnote{As opposed to the genetic algorithms used in \cite{Dutrieux:2021wll} for the optimization task.} \cite{DBLP:journals/corr/KingmaB14} to minimize the loss function, here chosen to be the \emph{Mean Squared Error} (MSE):
\begin{equation}
MSE = \frac{1}{N_{sample}} \sum_{i=1}^{N_{sample}} \left( \widehat{H}\left(x_{i},\xi_{i}\right) -H\left(x_{i},\xi_{i}\right)\right)^2.
\end{equation}
A \textit{Dropout} regularization is implemented \cite{JMLR:v15:srivastava14a}, where hidden neurons are randomly turned off with a fixed probability rate. It is worth noticing that the ill-posedness of a problem like the Radon transform inversion has been commonly treated using a \textit{Tikhonov} regularization. It has been shown, however, that these two approaches are essentially equivalent \cite{Bishop1995, WagerSidaLiang2013}, the dropout regularization being more effective for deeper networks.

The reader can note that the number of neurons and the choice of activation function play here a similar role compared to the number of cells discretizing space and the choice of interpolating polynomials, in the FEM approach. Yet, one can expect that the convergence and smoothness properties might be different between the two approaches.

\subsubsection{Algebraic model}\label{sec:ANNalg}

\begin{figure}[t]
\begin{subfigure}{.5\textwidth}
  \centering
  \includegraphics[width=0.95\linewidth]{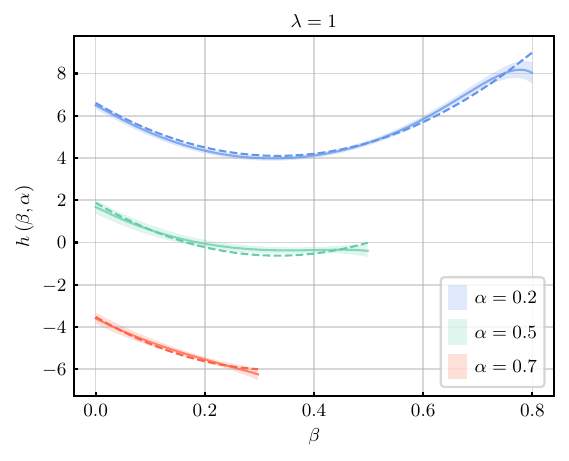}
\end{subfigure}%
\begin{subfigure}{.5\textwidth}
  \centering
  \includegraphics[width=0.95\linewidth]{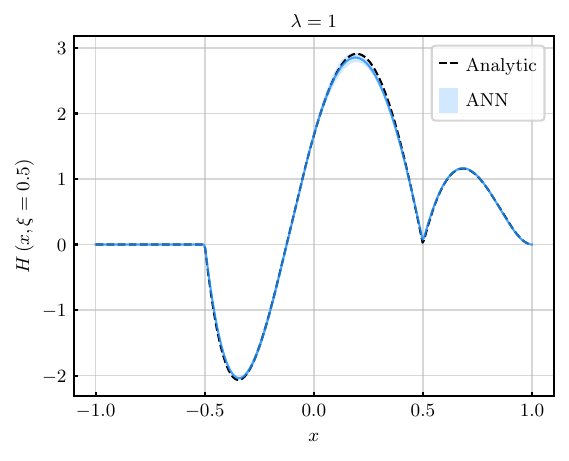}
\end{subfigure}
\caption{\small \textsc{(ANNs) Algebraic model} -- The left panel shows the DD $h_{\textrm{ANN}}(\beta, \alpha)$ obtained through the procedure described in Sec.~\ref{sec:ANNs} (solid lines), displayed as a function of the $\beta$ variable for three different values of $\alpha=\{0.2, 0.5, 0.7\}$ ($0\leq \beta \leq 1-\alpha$). For comparison, the exact analytical result of Eq.~\eqref{eq:alg_dd} is drawn through dashed lines. The sampling of the DGLAP region corresponds to $\xi\in\left[0,\lambda x\right]$ with $\lambda=1$ (the entire domain). The right panel displays the GPD $H\left(x,\xi\right)$ at the illustration value $\xi=0.5$, obtained as the Radon transform of the DD shown in the left, compared to the analytical expression, Eqs.~\eqref{eq:alg_dglap}~and~\eqref{eq:alg_erbl}.}
\label{fig:alg_l1}
\end{figure}

\begin{figure}[b]
\begin{subfigure}{.5\textwidth}
  \centering
  \includegraphics[width=0.95\linewidth]{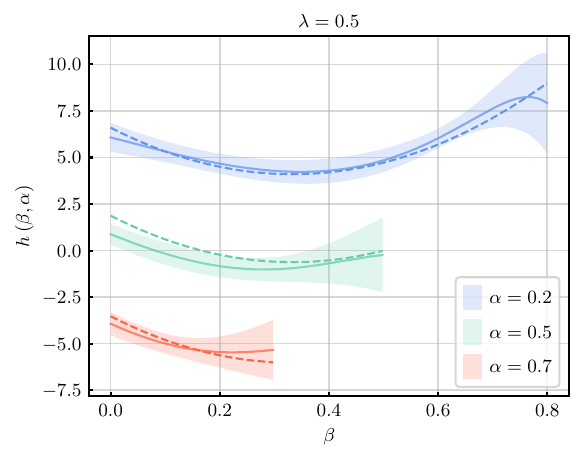}
\end{subfigure}%
\begin{subfigure}{.5\textwidth}
  \centering
  \includegraphics[width=0.95\linewidth]{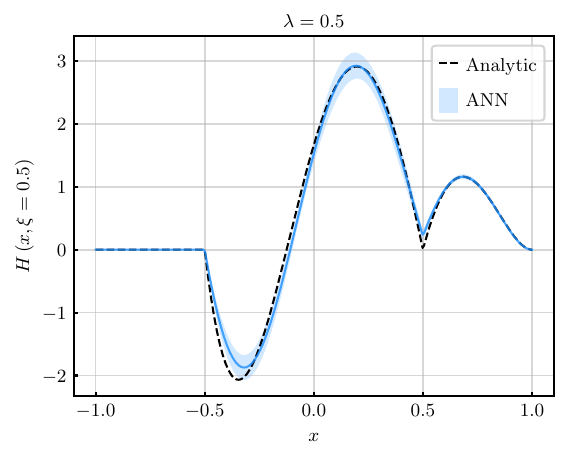}
\end{subfigure} 
\caption{\small \textsc{(ANNs) Algebraic model} -- The same of Fig.\,\ref{fig:alg_l1} with a DGLAP sampling given by $\xi\in\left[0,\lambda x\right]$ and $\lambda = 0.5$.}
\label{fig:alg_l05}
\end{figure}

In the case of the algebraic model, the ANN has been designed with $N=100$ hidden neurons, tuned by trial and error, and trained with a sample of $N_{sample} = 10^4$ GPD data points. Without loss of generality, both entries $x$ and $\xi$ are considered positive: The $x$ values are generated within the interval $\left[0,1\right]$, which is the $x$ domain for the quark sector in the DGLAP region, and the associated $\xi$ may remain also positive while covering the entire DGLAP region owing to the parity of the GPD. Still, for the inversion of the Radon transform \eqref{eq:RTP}, the subset $\mathcal{O}_{\lambda}$ in Eq.\,\eqref{eq:lambda_region} is fixed by restricting the skewness parameter to the interval $[0, \lambda x]$, with $0<\lambda\leq 1$, where the covariant extension is formally achievable, as discussed in Sec.~\ref{sec:RTInv-th}. 

However, despite the support theorem guarantees the uniqueness of the inverted Radon transform from $\mathcal{O}_\lambda$ with any $\lambda>0$; in practice, the noise in the obtained DD increases as the range of the skewness parameter becomes smaller. This is exhibited in the left panels of Figs. \ref{fig:alg_l1} and \ref{fig:alg_l05}, where three outcomes for the DD function are plotted, corresponding to the values of $\lambda = \{1, 0.5, 0.2 \}$. The values of $h(\beta, \alpha)$ are shown as functions of the $\beta$ variable for three different values of $\alpha$. The error bands associated with the ANN results correspond to a standard deviation from the mean value estimated by training the network from scratch on 50 independent trials (replicas).

The errors increase as $\lambda$ decreases, in particular closer to the boundary $\beta=1-\alpha$, where the DD is discontinuous. 
These somewhat large deviations of the numerical predictions from the exact analytical expression become milder, however, once the DD is integrated over the lines in order to get the GPD predictions. This is shown in the right panels of Figs.\,\ref{fig:alg_l1} and \ref{fig:alg_l05}, where the results displayed correspond to the particular value of $\xi =0.5$, chosen for illustrative purposes. This is noteworthy especially because the main goal of the current application of the support theorem is the GPD reconstruction; namely, the extension of GPD's knowledge from a restricted DGLAP domain to that on its entire support, particularly the full ERBL region. Although achieved by inferring the functional form of the DD, it is actually its Radon transform to be of primary interest here.

The kinematic restriction for the GPD data of the training set $\mathcal{O}_\lambda$ has been pushed down to a value as low as $\lambda=0.2$ in FEM. In this case, the quality of the results for both the inverted DD and the derived GPD is still good for $\lambda=0.5$, as it is shown in Fig.\,\ref{fig:alg_l05}, but it becomes strongly degraded for lower values of $\lambda$, these results not being then displayed. The main reason for this will be discussed below and comes as a drawback of the GPD behavior for the algebraic model approached with ANNs, presumably requiring further optimization. The GK model, in exchange, will be seen to admit, also  for $\lambda=0.2$, a very reliable GPD from a DD approximated with ANNs.

\subsubsection{Goloskokov-Kroll model}\label{sec:ANNgk}

In the case of the GK model, as shown in appendix \ref{app:GK}, the DD function $h_{\textrm{GK}}(\beta, \alpha)$ relies on a profile function $\pi_N(\beta;\alpha)$ as given by Eq.\,\eqref{eq:RDDA}, which fulfills the normalization condition\,\eqref{eq:norm-RDDA}. In the aim of making easier the explicit implementation of this condition, we have chosen this profile to be approximated by $\pi_{\textrm{ANN}}$ as it results from the ANN output $o^{(2)}$ (see Eq.~\eqref{eq:ANNform}) properly normalized. Thus, the GK DD is accordingly approximated by 

\begin{figure}[b]
\begin{subfigure}{.5\textwidth}
  \centering
  \includegraphics[width=0.95\linewidth]{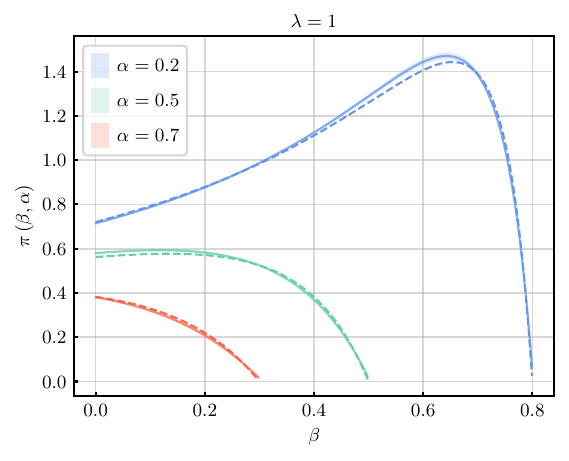}
\end{subfigure}%
\begin{subfigure}{.5\textwidth}
  \centering
  \includegraphics[width=0.95\linewidth]{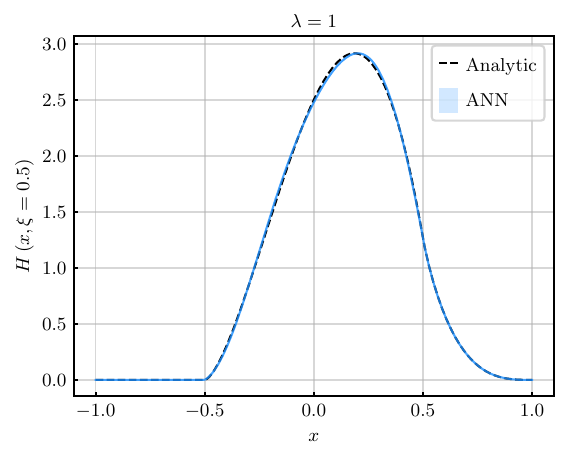}
\end{subfigure}
\caption{\small \textsc{(ANNs) GK model} -- The left panel shows the profile function, $\pi_{\textrm{ANN}}(\beta,\alpha)$ (solid lines), the direct ANN output which the DD follows immediately from as given in Eq.\,\eqref{eq:DD_gk_ANN}, obtained through the procedure described in Sec.~\ref{sec:ANNs} and displayed as a function of the $\beta$ variable for three different values of $\alpha=\{0.2, 0.5, 0.7\}$ ($0\leq \beta \leq 1-\alpha$). It compares strikingly well with the exact expressions, Eq.~\eqref{eq:pRDDA} with $N=1$(dashed lines). The sampling of the DGLAP region corresponds to $\xi\in\left[0,\lambda x\right]$ with $\lambda=1$ (the entire domain). The right plot displays the GPD $H\left(x,\xi\right)$ at the illustration value $\xi=0.5$, obtained as the Radon transform of the DD shown in the left, compared to the analytical expression, Eqs.~\eqref{eq:gk_dglap}~and~\eqref{eq:gk_erbl}.}
\label{fig:gk_val_l1}
\end{figure}

\begin{equation}\label{eq:DD_gk_ANN}
h_{\textrm{ANN}}(\beta, \alpha) = f(\beta;\mu) \, \frac{o^{(2)}(\beta, \alpha)}{\int_{-1+|\beta|}^{1-|\beta|}d\alpha'\, o^{(2)}(\beta,\alpha')}\;;
\end{equation}
where $f(\beta,\mu)$ is the parton distribution function given by Eq.\,\eqref{eq:cteqFit}. Notice that in Ref.\,\cite{Dutrieux:2021wll}, in addition to implementing this normalization condition, the ANN expression \eqref{eq:ANNform} was modified in order to impose the vanishing of the output along the border $\alpha = 1-\beta$, a property which is indeed satisfied by the $h_{\textrm{GK}}\left(\beta,\alpha\right)$ in \eqref{eq:dd_gk}. We decided however not to implement this constraint explicitly and let the network learn it by itself, considering that the vanishing at the boundary is a particular feature of this model, unlike the normalization condition and the parity in the $\alpha$ variable, which are properties that come from first principles. This decision comes at the cost of a much slower convergence of the optimization algorithm, due to the larger functional space that has to be explored during the training phase.

\begin{figure}[b]
\begin{subfigure}{.5\textwidth}
\begin{subfigure}{\textwidth}
  \centering
  \includegraphics[width=0.95\linewidth]{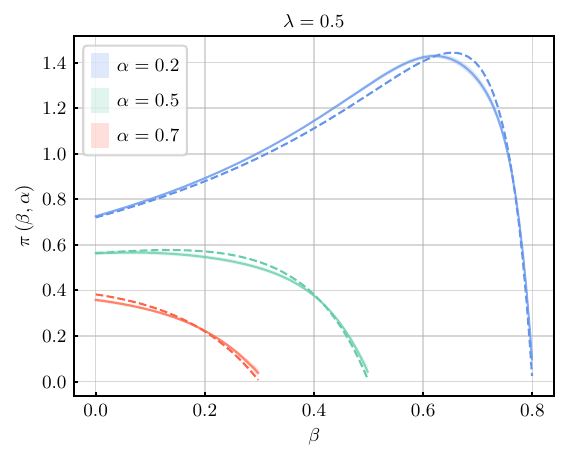}
\end{subfigure}
\begin{subfigure}{\textwidth}
  \centering
  \includegraphics[width=0.95\linewidth]{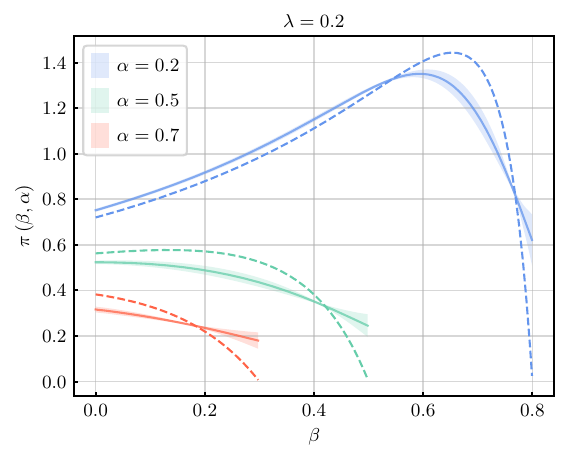}
\end{subfigure}
\end{subfigure}%
\begin{subfigure}{.5\textwidth}
\begin{subfigure}{\textwidth}
  \centering
  \includegraphics[width=0.95\linewidth]{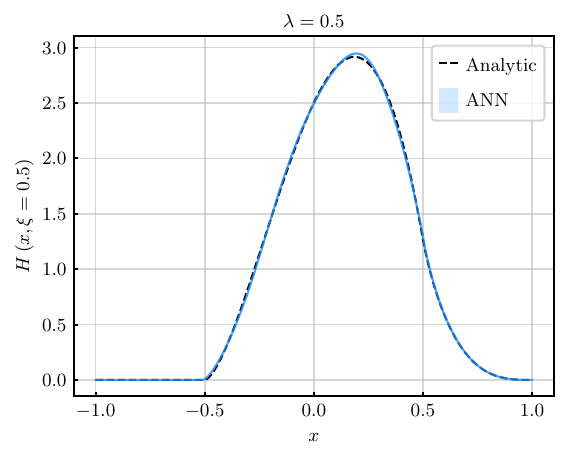}
\end{subfigure}
\begin{subfigure}{\textwidth}
  \centering
  \includegraphics[width=0.95\linewidth]{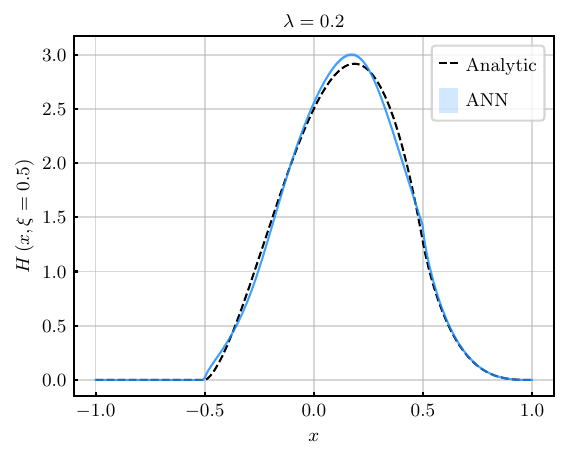}
\end{subfigure}
\end{subfigure}%
\caption{\small \textsc{(ANNs) GK model} -- The same of Fig.\,\ref{fig:gk_val_l1} with a DGLAP sampling given by $\xi\in\left[0,\lambda x\right]$, and $\lambda = 0.5$ (top), $0.2$ (bottom).}
\label{fig:gk_val_l05}
\end{figure}

As opposed to the previous example, the DD function that the network has to learn is now everywhere continuous, since it vanishes at the boundary of its support. This property helps the ANN to better converge to the exact solution, yielding more precise results, as it can be appreciated in Figs.~\ref{fig:gk_val_l1} and \ref{fig:gk_val_l05}. There, in the left panels, we have chosen to represent the profile function, $\pi_{\textrm{ANN}}(\beta,\alpha)$,  the direct output from ANNs which the DD follows immediately from as given in Eq.\,\eqref{eq:DD_gk_ANN}. 
Furthermore, a smaller network was used, with $N=25$ hidden neurons, and a smaller training data sampling of $N_{sample}=5\times 10^3$  GPD values generated in the DGLAP region using Eq.\,\eqref{eq:gk_dglap}, derived from Eqs.\,(\ref{eq:cteqFit},\ref{eq:dd_gk}). In the case of a kinematic restriction given by $\lambda=0.2$, the prediction for the inverse function is rather imprecise, although the deviation from the exact result of its Radon transform is not substantial (see the bottom panels in Fig. \ref{fig:gk_val_l05}). 

As stated above, we do not aim at a careful and conclusive comparison of the two different approaches herein followed to invert the Radon transform; namely, ANNs and FEM. Both of them have been implemented instead to avoid drawing biased conclusions about our methodology that may stem for the suitability of a particular implementation to the models chosen as an illustration. To this aim, we have independently tuned the setups in both ANN and FEM approaches such that they deliver good results when they compare to the exact ones. In particular, we have presented results in which the ratio of ANN to FEM input data is roughly $100$. 

Notwithstanding, some exploratory studies in the context of approximating the GK model through the ANN approach evidence that still acceptable results, while poorer, could be achieved by reducing the number of input data to $250$. However, similar studies around the algebraic model seem to point out an issue arising from the use of non-polynomial activation functions, requiring a large number of neurons to reproduce the polynomial behavior of the algebraic DD. The use of extra neurons being associated to the need of more training data, this observation suggests a limitation of the ANN-based method. One could instead consider different activation functions, better adapted to a polynomial DD, but this would be an \emph{ad-hoc} improvement relying on \emph{prior} knowledge of the DD. On the other hand, techniques of data augmentation which are commonly used in machine learning applications could be also implemented. Indeed, we have triggered an analysis with $250$ data points and used standard interpolation procedures to generate enough additional data to invert the Radon transform for the algebraic model GPD, thus getting results similar to those obtained with FEM. Nevertheless, this analysis neglects the impact of errors and noise, which will become also strongly amplified by the data augmentation. A careful and systematic analysis, considering noise and ANNs optimization would be needed to perform a reliable comparison with FEM. This is anyhow out of the scope of the present work.

\section{Summary and conclusions}

In this work we have demonstrated that, at fixed Mandelstam $t$, Lorentz covariance as encoded through the renowned polynomiality imposes a strong enough constraint so that the sole knowledge of GPDss at zero- and very-low skewness is enough to have them determined over their entire kinematic range. The formal proof of this result relies entirely on a theorem by Boman and Todd-Quinto \cite{boman:1987rad}, which we have revisited here and made it more transparent for the field of GPDs by its formulation in the appropriate language. 

We have thus concluded that the knowledge of GPDs for $x\in\left[-1,1\right]$ and $\xi\in\left[0,\lambda x\right]$ is enough to fully constraint them over their entire $\left(x,\xi\right)$ range. Furthermore, this idea has been implemented in practice through two different algorithms: One based on ANNs, and a second one using a more traditional strategy using FEM. Both approaches have been tested against standard models for GPDs, showing that accurate reconstructions shall be obtained in practice with input information as restricted as to $\sim 20~\%$ of the DGLAP region. Given the increasing interest in the study of GPDs, we believe this approach to be valuable in complementing the limited kinematic range over which we expect outcomes to be obtained from Lattice QCD simulations (limited by Ioffe-time and hadron momentum accessible, see \textit{e.g.}\cite{Cichy:2023dgk}) and experimental data. So far, looking towards a practical application of this methodology, the FEM-strategy seems to perform better than the ANN one: The number of inputs required by the ANN approach is ten times larger than the one needed in the FEM approach. The reason is that in the former case, the required information roughly linearly scales with the size of the network. Nevertheless, optimization of the ANN approach can still be achieved, as well as the use of data augmentation techniques implemented. This work being intended at drawing the attention to this novel technique and not to a refined kinematic completion of actual data, added to the lack of reliable GPD extractions, we defer such a comparative study, optimization of the numerical techniques and testing of robustness against noise to a future work.

\section*{Acknowledgments}

JMMC thank V. Bertone, B. Blossier, M. Riberdy and T. San Jos\'e for valuable discussions and comments. The work from CM, JMMC and HM has been supported by the GLUODYNAMICS project funded by ``P2IO LabEx (ANR-10-LABX-0038)'' in the framework of Investissements d’Avenir (ANR-11-IDEX-0003-01), managed by the Agence Nationale de la Recherche (ANR), France; and by the European Union's Horizon 2020 research and innovation program under grant agreement STRONG 2020 – No. 824093. PDO acknowledges financial support from ``Junta de Andaluc\'ia'' through "Programa Operativo FEDER de Andalucía 2014-2020 (PAIDI)" under the project P20\_00764. The work of PDO, FDS, JR-Q and JS was supported by the ``Spanish Ministerio de Ciencia e Innovaci\'on (MICINN)'' through grants PID2019-107844-GB-C22 and PID2022-140440-NB-C22.

\appendix

\section{Models for generalized parton distributions}

In this work we have illustrated our methodology using two different, well known models for GPDss: $(a)$ The \textit{algebraic model} \cite{Mezrag:2016hnp, Chouika:2017rzs, Chouika:2017dhe} and $(b)$ the \textit{Goloskokov-Kroll model} \cite{Goloskokov:2005sd,Goloskokov:2006hr,Goloskokov:2007nt,Goloskokov:2009ia}. There are two main reasons for this choice. On the one hand, the so called algebraic model is currently being exploited for the description of pion's structure (\textit{e.g.} \cite{Chavez:2021llq, Chavez:2021koz}) while the Goloskokov-Kroll model is popular in the study of nucleon GPDs, see \textit{e.g.} \cite{Moutarde:2018kwr,Kumericki:2016ehc} and references therein. We thus find enlightening to benchmark our approach on the (perhaps) two most paradigmatic scenarios found in the study of hadron structure, pions and nucleons, thus triggering our interest into these two models. A second argument follows from a more technical perspective: If the DD of the algebraic model is a smooth function of its kinematic variables, Sec.~\ref{app:Alg}, that of the GK model shows an integrable divergence at $\beta\rightarrow 0$, Sec.~\ref{app:GK}. Therefore we find appropriate to benchmark our implementations in these two radically different scenarios where numerical accuracy is expected to be challenged in two different manners. 

This being said, the two models could have been employed to test our algorithms without any regard the underlying physics. In fact, we are here simply testing our way of solving the inverse Radon transform problem. However, for the sake of self-consistency, we believe worthwhile briefly reviewing their foundations.

\subsection{Algebraic model}\label{app:Alg}

Capitalizing on a long effort towards the development of models for parton distributions respecting all of the QCD's basic properties \cite{Mezrag:2013mya,Mezrag:2014jka,Mezrag:2014tva,Mezrag:2015mka,Mezrag:2016hnp,Chouika:2017dhe}, the \textit{algebraic model} for the GPD of valence-quarks in a pion was first presented in Ref.~\cite{Chouika:2017rzs}. There the authors relied on the overlap representation of GPDs \cite{Diehl:2000xz, Diehl:2003ny} to model the DGLAP region in a way consistent with their positivity property \cite{Pire:1998nw,Pobylitsa:2002gw}. To that end, the essential ingredient is a parametrization of the hadron's light-front wave function (LFWF) $\Psi\left(x,\bm{k}_{\perp};\mu\right)$; which for a two-body system reads \cite{Diehl:2000xz}
\begin{equation}\label{eq:GPDoverlap}
    \left.H\left(x,\xi;\mu\right)\right|_{\textrm{DGLAP}}=\sum_{\sigma}\int\frac{d^{2}\bm{k}_{\perp}}{16\pi^{3}}\Psi^{\ast}_{\sigma}\left(\frac{x-\xi}{1-x},\bm{k}_{\perp}+\frac{1-x}{1-\xi}\frac{\bm{\Delta}_{\perp}}{2};\mu\right)\Psi_{\sigma}\left(\frac{x+\xi}{1+\xi},\bm{k}_{\perp}-\frac{1-x}{1+\xi}\frac{\bm{\Delta}_{\perp}}{2};\mu\right),
\end{equation}
$\sigma$ being the possible quark-helicity combinations, $\bm{k}_{\perp}$ the average momentum of the quark pair along the transverse direction, as defined by the hadron's (light-cone) momentum, $p$; and $\bm{\Delta}_{\perp}$ the transverse components of the four-momentum transfer between quarks.

The light-front wave function giving rise to the algebraic model was obtained in Euclidean space from an appropriate projection of the pion's Bethe-Salpeter wave function \cite{Chang:2013pq}, which was in turn constructed using a Nakanishi representation for the Bethe-Salpeter amplitude \cite{Nakanishi:1963zz,Nakanishi:1969ph}
\begin{equation}
    \chi\left(k,p\right)=i\mathcal{N}\gamma_{5}\int_{0}^{\infty}dw\int_{-1}^{1}dz\frac{\rho\left(w,z\right)M^{2}}{\left(k-\frac{1-z}{2}p\right)^{2}+M^{2}+w},\qquad\textrm{with }\rho\left(w,z\right)=\delta\left(w\right)\left(1-z^{2}\right).
\end{equation}
where $M$ is a mass-scale, supplemented with an \textit{Ansatz} of the form  
\begin{equation}
    S\left(k\right)=\left(-i\gamma\cdot k+M\right)/\left(k^{2}+M^{2}\right),
\end{equation}
for the quark-propagator.

Such a parametrization allows for a fully algebraic treatment leading to the evaluation of the two possible light-front wave functions in closed form \cite{Mezrag:2016hnp}, thus the denomination of the resulting GPD model as ``algebraic''. Feeding the overlap representation Eq.~\eqref{eq:GPDoverlap} with those LFWFs yields 
\begin{equation}\label{eq:alg_dglap}
    \left.H\left(x,\xi;\mu\right)\right|_{\textrm{DGLAP}}=30\frac{\left(1-x\right)^{2}\left(x^{2}-\xi^{2}\right)}{\left(1-\xi^{2}\right)^{2}}.
\end{equation}

Here from, the companion ERBL region was constructed in \cite{Chouika:2017rzs} following the dictations of Lorentz covariance as captured by the renowned polynomiality property and implemented through the (inverse) Radon transform \cite{Teryaev:2001qm, Chouika:2017dhe}. As a result, the underlying DD is found \cite{Chouika:2017rzs}
\begin{equation}\label{eq:alg_dd}
    h_{\textrm{Alg.}}\left(\beta,\alpha;\mu\right)=\frac{15}{2}\left[1-3\left(\alpha^{2}-\beta^{2}\right)-2\beta\right].
\end{equation}
while the ERBL GPD turns out to be
\begin{equation}\label{eq:alg_erbl}
    \left.H\left(x,\xi;\mu\right)\right|_{\textrm{ERBL}}=\frac{15}{2}\frac{\left(1-x\right)\left(x^{2}-\xi^{2}\right)}{\xi^{3}\left(1+\xi\right)^{2}}\left(x+2x\xi+\xi^{2}\right).
\end{equation}

As a final remark notice that no explicit reference has been made to the scale $\mu$ in the presentation of this model. Its setting is implicit in the expression Eq.~\eqref{eq:GPDoverlap}, where a meson is being represented by two-body LFWFs. A discussion about this subject is definitely outside the scope of the present work, the interested reader can find more information in \textit{e.g.} \cite{MorgadoChavez:2022men} and references therein.  

\subsection{Goloskokov-Kroll model}\label{app:GK}

The Goloskokov-Kroll model (GK) \cite{Goloskokov:2005sd,Goloskokov:2006hr,Goloskokov:2007nt,Goloskokov:2009ia} is a popular phenomenological parametrization, originally tailored for the description of deeply virtual meson production (DVMP) but which has also proved to yield good agreement in a leading order description of deeply virtual Compton scattering data\footnote{For the sake of completeness one must also point out that such agreement was argued in \cite{Kumericki:2015lhb} to be only casual.} \cite{Kroll:2012sm}. It is built on top of \textit{Radyushkin's double distribution Ansatz} (RDDA) which in turn relays on two simple ideas \cite{Musatov:1999xp}:
\begin{itemize}
    \item The profile of a GPD along the $x$-direction is basically determined by that of $f\left(x;\mu\right)\equiv H\left(x,\xi=0;\mu\right)$, \textit{i.e.} the parton distribution function.
    \item The shape of a GPD along the $\xi$-direction characterizes the spread of parton momentum induced by momentum transfer.
\end{itemize}

A simple way to combine these two assumptions is to model GPDs through DDs of the form:
\begin{equation}\label{eq:RDDA}
    h_{\textrm{RDDA}}^{\left(N\right)}\left(\beta,\alpha;\mu\right)=f\left(\beta;\mu\right)\pi_{N}\left(\beta,\alpha\right) \;;
\end{equation}
where $\pi_{N}\left(\beta,\alpha\right)$ is a profile function. From that point on Eq.~\eqref{eq:RT0} gives
\begin{equation}
    H\left(x,\xi=0;\mu\right)=\iint_{\Omega}d\beta d\alpha\delta\left(x-\beta\right)f\left(\beta;\mu\right)\pi_{N}\left(\beta,\alpha\right)=f\left(x;\mu\right)\int_{-1+\left|x\right|}^{1-\left|x\right|}\pi_{N}\left(x,\alpha\right) \;,
\end{equation}
which produces the desired forward limit if the profile function is normalized as
\begin{equation}\label{eq:norm-RDDA}
    \int_{-1+\left|x\right|}^{1-\left|x\right|}d\alpha\pi_{N}\left(x,\alpha\right)=1 \;.
\end{equation}

A simple way to fulfill such requirement is to employ the profile function suggested in \cite{Musatov:1999xp}
\begin{equation}\label{eq:pRDDA}
    \pi_{N}\left(\beta,\alpha\right)=\frac{\Gamma\left(2N+2\right)}{2^{2N+1}\Gamma^{2}\left(N+1\right)}\frac{\left[\left(1-\left|\beta\right|\right)^{2}-\alpha^{2}\right]^{N}}{\left(1-\left|\beta\right|\right)^{2N+1}} \;;
\end{equation}
which satisfies the condition Eq.~\eqref{eq:norm-RDDA} $\forall~N~|~\textrm{Re}\left(N\right)>-1$. Strikingly, the $\alpha$-dependence of such profile function is entirely controlled by a single parameter, $N$, producing a rather inflexible modeling\footnote{In fact, a rapid convergence of the models thus produced through different choices of $N$ was demonstrated in \cite{Mezrag:2013mya,Mezrag:2015mka}.}.

In \cite{Goloskokov:2005sd,Goloskokov:2006hr,Goloskokov:2007nt,Goloskokov:2009ia} this approach is followed to build the renowned Goloskokov-Kroll model. There, the valence-quark GPD in nucleons is designed choosing $N=1$ in the profile function $\pi_{N}\left(\beta,\alpha\right)$, which reproduces the expected asymptotic behavior for the quark distribution amplitude \cite{Goloskokov:2005sd}; and employing a phenomenological \textit{Ansatz} for the parton distribution function,
\begin{equation}\label{eq:cteqFit}
    f\left(\beta;\mu\right)=\beta^{-\delta}\left(1-\beta\right)^{3}\sum_{j=0}^{2}c_{j}\left(\mu\right)\beta^{j/2}
\end{equation}
with $\delta$- and $c_{j}$-parameters determined from a fit to the CTEQ6m PDF \cite{Goloskokov:2006hr, Pumplin:2002vw} (see Tab.~\ref{tab:GK-params}).

\begin{table}[t]
    \centering
    \begin{tabular}{c|c}\hline\hline
       $\delta$  & $0.48$        \\
       $c_{0}\left(\mu\right)$   & $1.52+0.248\log(\mu^{2}/\mu_{0}^{2})$\\
       $c_{1}\left(\mu\right)$   & $2.88-0.940\log(\mu^{2}/\mu_{0}^{2})$\\
       $c_{2}\left(\mu\right)$   & $-0.095\log(\mu^{2}/\mu_{0}^{2})$    \\\hline\hline
    \end{tabular}
    \caption{Parameters of the \textit{Ansatz} Eq.~\eqref{eq:cteqFit} fitting the CTEQ6m PDF \cite{Pumplin:2002vw} in the range $10^{-2}\leq\beta\leq 0.5$ and $\mu^{2}_{0}\equiv 4~\textrm{GeV}^{2}\leq\mu^{2}\leq 40~\textrm{GeV}^{2}$. Taken from \cite{Goloskokov:2006hr}.}
    \label{tab:GK-params}
\end{table}

Combining the parametrization above with Radyushkin's DD \textit{Ansatz} generates the Goloskokov-Kroll model for the DD of valence-quarks within nucleons 
\begin{equation}\label{eq:dd_gk}
    h_{\textrm{GK}}\left(\beta,\alpha;\mu\right)\equiv h_{\textrm{RDDA}}^{\left(1\right)}\left(\beta,\alpha;\mu\right)=\frac{3}{4}\left(\left(1-\beta\right)^{2}-\alpha^{2}\right)\beta^{-\delta}\sum_{j=0}^{2}c_{j}\left(\mu\right)\beta^{j/2}.
\end{equation}

For the sake of simplicity, in this work we shall deal with the present model at $\mu^{2}=\mu_{0}^{2}=4~\textrm{GeV}^{2}$ which leaves us with the GPD
\begin{equation}\label{eq:gk_dglap}
\begin{array}{rcl}
\displaystyle \left.H\left(x,\xi;\mu\right)\right|_{\textrm{DGLAP}} & \displaystyle = & \displaystyle \frac{3}{4\xi^{3}}\sum_{j=0}^{2}\frac{c_{j}}{\left(1-\xi^{2}\right)^{a+1}}\times\\
\\
&\displaystyle \times & \displaystyle \left\lbrace\frac{\xi^{2}-x^{2}}{a+1}\left[\left(x+\xi\right)^{a+1}\left(1-\xi\right)^{a+1}-\left(x-\xi\right)^{a+1}\left(1+\xi\right)^{a+1}\right]\right.  \\
\\
& \displaystyle & \displaystyle -2\frac{\xi^{2}-x}{\left(a+2\right)\left(1-\xi^{2}\right)}\left[\left(x+\xi\right)^{a+2}\left(1-\xi\right)^{a+2}-\left(x-\xi\right)^{a+2}\left(1+\xi\right)^{a+2}\right] \\
\\
& \displaystyle & \displaystyle +\left.\frac{\xi^{2}-1}{\left(a+3\right)\left(1-\xi^{2}\right)^{2}}\left[\left(x+\xi\right)^{a+3}\left(1-\xi\right)^{a+3}-\left(x-\xi\right)^{a+3}\left(1+\xi\right)^{a+3}\right]\right\rbrace,\\
\end{array}
\end{equation}
within the DGLAP region, and
\begin{equation}\label{eq:gk_erbl}
    \displaystyle \left.H\left(x,\xi;\mu\right)\right|_{\textrm{ERBL}} = \frac{3}{4\xi^{3}}\sum_{j=0}^{2}c_{j}\left(\frac{x+\xi}{1+\xi}\right)^{a+1}\left\lbrace\frac{\xi^{2}-x^{2}}{a+1}+\frac{x+\xi}{1+\xi}\left[\frac{\left(x+\xi\right)\left(\xi-1\right)}{a+3}-2\frac{\xi^{2}-x}{a+2}\right]\right\rbrace,
\end{equation}
in the ERBL region, where $a\equiv j/2-\delta$.

\bibliographystyle{unsrt}
\bibliography{Bibliography.bib}

\begin{thebibliography}{10}

\bibitem{Mueller:1998fv}
D.~Mueller, D.~Robaschik, B.~Geyer, F.~M. Dittes, and J.~Ho\v{r}e\v{j}si.
\newblock {Wave functions, evolution equations and evolution kernels from light
  ray operators of QCD}.
\newblock {\em Fortsch.Phys.}, 42:101--141, 1994.

\bibitem{Ji:1996nm}
X.~Ji.
\newblock {Deeply virtual Compton scattering}.
\newblock {\em Phys.Rev.}, D55:7114--7125, 1997.

\bibitem{Radyushkin:1997ki}
A.V. Radyushkin.
\newblock {Nonforward parton distributions}.
\newblock {\em Phys.Rev.}, D56:5524--5557, 1997.

\bibitem{Burkardt:2000za}
M.~Burkardt.
\newblock {Impact parameter dependent parton distributions and off forward
  parton distributions for zeta ---> 0}.
\newblock {\em Phys. Rev.}, D62:071503, 2000.
\newblock [Erratum: Phys. Rev.D66,119903(2002)].

\bibitem{Diehl:2002he}
M.~Diehl.
\newblock {Generalized parton distributions in impact parameter space}.
\newblock {\em Eur.Phys.J.}, C25:223--232, 2002.

\bibitem{Ji:1996ek}
X.~Ji.
\newblock {Gauge-Invariant Decomposition of Nucleon Spin}.
\newblock {\em Phys. Rev. Lett.}, 78:610--613, 1997.

\bibitem{Polyakov:2002yz}
M.~V. Polyakov.
\newblock {Generalized parton distributions and strong forces inside nucleons
  and nuclei}.
\newblock {\em Phys. Lett.}, B555:57--62, 2003.

\bibitem{Kumericki:2015lhb}
K.~Kumerički and D.~Müller.
\newblock {Description and interpretation of DVCS measurements}.
\newblock {\em EPJ Web Conf.}, 112:01012, 2016.

\bibitem{Moutarde:2018kwr}
H.~Moutarde, P.~Sznajder, and J.~Wagner.
\newblock {Border and skewness functions from a leading order fit to DVCS
  data}.
\newblock {\em Eur. Phys. J.}, C78(11):890, 2018.

\bibitem{Bertone:2021yyz}
V.~Bertone, H.~Dutrieux, C.~Mezrag, H.~Moutarde, and P.~Sznajder.
\newblock {The deconvolution problem of deeply virtual Compton scattering}.
\newblock {\em Phys. Rev. D}, 103(11):114019, 4 2021.

\bibitem{Moffat:2023svr}
E.~Moffat, A.~Freese, I.~Clo\"et, T.~Donohoe, L.~Gamberg, W.~Melnitchouk,
  A.~Metz, A.~Prokudin, and N.~Sato.
\newblock {Shedding light on shadow generalized parton distributions}.
\newblock {\em Phys. Rev. D}, 108(3):036027, 2023.

\bibitem{Boussarie:2016qop}
R.~Boussarie, B.~Pire, L.~Szymanowski, and S.~Wallon.
\newblock {Exclusive photoproduction of a $\gamma\,\rho$ pair with a large
  invariant mass}.
\newblock {\em JHEP}, 02:054, 2017.
\newblock [Erratum: JHEP 10, 029 (2018)].

\bibitem{Duplancic:2023kwe}
G.~Duplan\v{c}i\'c, S.~Nabeebaccus, K.~Passek-Kumeri\v{c}ki, B.~Pire,
  L.~Szymanowski, and S.~Wallon.
\newblock {Probing chiral-even and chiral-odd leading twist quark generalised
  parton distributions through the exclusive photoproduction of a $ \gamma \rho
  $ pair}.
\newblock 2 2023.

\bibitem{Grocholski:2021man}
O.~Grocholski, B.~Pire, P.~Sznajder, L.~Szymanowski, and J.~Wagner.
\newblock {Collinear factorization of diphoton photoproduction at next to
  leading order}.
\newblock {\em Phys. Rev. D}, 104(11):114006, 2021.

\bibitem{Qiu:2022bpq}
J.-W. Qiu and Z.~Yu.
\newblock {Exclusive production of a pair of high transverse momentum photons
  in pion-nucleon collisions for extracting generalized parton distributions}.
\newblock {\em JHEP}, 08:103, 2022.

\bibitem{Qiu:2023mrm}
J.-W. Qiu and Z.~Yu.
\newblock {Extraction of the Parton Momentum-Fraction Dependence of Generalized
  Parton Distributions from Exclusive Photoproduction}.
\newblock {\em Phys. Rev. Lett.}, 131(16):161902, 2023.

\bibitem{Ji:1998pc}
X.~Ji.
\newblock {Off forward parton distributions}.
\newblock {\em J.Phys.}, G24:1181--1205, 1998.

\bibitem{Radyushkin:1998bz}
A.V. Radyushkin.
\newblock {Symmetries and structure of skewed and double distributions}.
\newblock {\em Phys.Lett.}, B449:81--88, 1999.

\bibitem{Radyushkin:1998es}
A.V. Radyushkin.
\newblock {Double distributions and evolution equations}.
\newblock {\em Phys.Rev.}, D59:014030, 1999.

\bibitem{Pire:1998nw}
B.~Pire, J.~Soffer, and O.~Teryaev.
\newblock {Positivity constraints for off - forward parton distributions}.
\newblock {\em Eur.Phys.J.}, C8:103--106, 1999.

\bibitem{Diehl:2000xz}
M.~Diehl, T.~Feldmann, R.~Jakob, and P.~Kroll.
\newblock {The Overlap representation of skewed quark and gluon distributions}.
\newblock {\em Nucl.Phys.}, B596:33--65, 2001.

\bibitem{Pobylitsa:2002gw}
P.V. Pobylitsa.
\newblock {Disentangling positivity constraints for generalized parton
  distributions}.
\newblock {\em Phys.Rev.}, D65:114015, 2002.

\bibitem{Chavez:2021llq}
J.~M. Morgado~Ch\'avez, V.~Bertone, F.~De~Soto, M.~Defurne, C.~Mezrag,
  H.~Moutarde, J.~Rodr\'\i{}guez-Quintero, and J.~Segovia.
\newblock {Pion generalized parton distributions: A path toward phenomenology}.
\newblock {\em Phys. Rev. D}, 105(9):094012, 2022.

\bibitem{Chouika:2017dhe}
N.~Chouika, C.~Mezrag, H.~Moutarde, and J.~Rodríguez-Quintero.
\newblock {Covariant Extension of the GPD overlap representation at low Fock
  states}.
\newblock {\em Eur. Phys. J.}, C77:906, 2017.

\bibitem{Chouika:2017rzs}
N.~Chouika, C.~Mezrag, H.~Moutarde, and J.~Rodríguez-Quintero.
\newblock {A Nakanishi-based model illustrating the covariant extension of the
  pion GPD overlap representation and its ambiguities}.
\newblock {\em Phys. Lett.}, B780:287--293, 2018.

\bibitem{Chavez:2021koz}
J.~M. Morgado~Ch\'avez, V.~Bertone, F.~De~Soto, M.~Defurne, C.~Mezrag,
  H.~Moutarde, J.~Rodr\'\i{}guez-Quintero, and J.~Segovia.
\newblock {Accessing the Pion 3D Structure at US and China Electron-Ion
  Colliders}.
\newblock {\em Phys. Rev. Lett.}, 128(20):202501, 2022.

\bibitem{Dutrieux:2023qnz}
H.~Dutrieux, M.~Winn, and V.~Bertone.
\newblock {Exclusive meets inclusive particle production at small Bjorken xB:
  How to relate exclusive measurements to PDFs based on evolution equations}.
\newblock {\em Phys. Rev. D}, 107(11):114019, 2023.

\bibitem{Riberdy:2023awf}
M.~J. Riberdy, H.~Dutrieux, C.~Mezrag, and P.~Sznajder.
\newblock {Combining lattice QCD and phenomenological inputs on generalised
  parton distributions at moderate skewness}.
\newblock 6 2023.

\bibitem{Karpie:2021pap}
J.~Karpie, K.~Orginos, A.~Radyushkin, and S.~Zafeiropoulos.
\newblock {The continuum and leading twist limits of parton distribution
  functions in lattice QCD}.
\newblock {\em JHEP}, 11:024, 2021.

\bibitem{Bhattacharya:2022aob}
S.~Bhattacharya, K.~Cichy, M.~Constantinou, J.~Dodson, X.~Gao, A.~Metz,
  S.~Mukherjee, A.~Scapellato, F.~Steffens, and Y.~Zhao.
\newblock {Generalized parton distributions from lattice QCD with asymmetric
  momentum transfer: Unpolarized quarks}.
\newblock {\em Phys. Rev. D}, 106(11):114512, 2022.

\bibitem{Bhattacharya:2023nmv}
S.~Bhattacharya, K.~Cichy, M.~Constantinou, J.~Dodson, A.~Metz, A.~Scapellato,
  and F.~Steffens.
\newblock {Chiral-even axial twist-3 GPDs of the proton from lattice QCD}.
\newblock {\em Phys. Rev. D}, 108(5):054501, 2023.

\bibitem{Goloskokov:2005sd}
S.V. Goloskokov and P.~Kroll.
\newblock {Vector meson electroproduction at small Bjorken-x and generalized
  parton distributions}.
\newblock {\em Eur.Phys.J.}, C42:281--301, 2005.

\bibitem{Polyakov:1999gs}
Maxim~V. Polyakov and C.~Weiss.
\newblock {Skewed and double distributions in pion and nucleon}.
\newblock {\em Phys.Rev.}, D60:114017, 1999.

\bibitem{Diehl:2003ny}
M.~Diehl.
\newblock {Generalized parton distributions}.
\newblock {\em Phys.Rept.}, 388:41--277, 2003.

\bibitem{Natterer:2001}
F.~Natterer and F.~Wübbeling.
\newblock {\em Mathematical Method in image Reconstruction}.
\newblock Monographs on Mathematical Modeling and Computation, Society for
  Industrial and Applied Mathematics, New York, 2001.

\bibitem{Hertle:1983}
A.~Hertle.
\newblock Continuity of the radon transform and its inverse on euclidean space.
\newblock {\em Mathematische Zeitschrift}, 184(2):165--192, 1983.

\bibitem{Chouika:2018mbk}
N.~Chouika.
\newblock {\em {Generalized Parton Distributions and their covariant extension:
  towards nucleon tomography}}.
\newblock PhD thesis, IRFU, Saclay, DPHN, 2018.

\bibitem{Teryaev:2001qm}
O.V. Teryaev.
\newblock {Crossing and radon tomography for generalized parton distributions}.
\newblock {\em Phys.Lett.}, B510:125--132, 2001.

\bibitem{Tiburzi:2004qr}
B.C. Tiburzi.
\newblock {Double distributions: Loose ends}.
\newblock {\em Phys.Rev.}, D70:057504, 2004.

\bibitem{Pobylitsa:2002vi}
P.V. Pobylitsa.
\newblock {Solution of polynomiality and positivity constraints on generalized
  parton distributions}.
\newblock {\em Phys.Rev.}, D67:034009, 2003.

\bibitem{boman:1987rad}
J.~Boman and E.~Todd-Quinto.
\newblock Support theorems for real-analytic radon transforms.
\newblock {\em Duke Math. J.}, 55(4):943--948, 12 1987.

\bibitem{Chouika:2017itz}
N.~Chouika, C.~Mezrag, H.~Moutarde, and J.~Rodríguez-Quintero.
\newblock {Concurrent approaches to Generalized Parton Distribution modeling:
  the pion’s case}.
\newblock {\em EPJ Web Conf.}, 137:05020, 2017.

\bibitem{MorgadoChavez:2022men}
J.~M. Morgado~Ch\'avez.
\newblock {\em {Generalized parton distributions of the pion: modeling,
  evolution and observable implications}}.
\newblock PhD thesis, Huelva U., 2022.

\bibitem{Goloskokov:2006hr}
S.V. Goloskokov and P.~Kroll.
\newblock {The Longitudinal cross-section of vector meson electroproduction}.
\newblock {\em Eur.Phys.J.}, C50:829--842, 2007.

\bibitem{Goloskokov:2007nt}
S.V. Goloskokov and P.~Kroll.
\newblock {The Role of the quark and gluon GPDs in hard vector-meson
  electroproduction}.
\newblock {\em Eur.Phys.J.}, C53:367--384, 2008.

\bibitem{Goloskokov:2009ia}
S.V. Goloskokov and P.~Kroll.
\newblock {An Attempt to understand exclusive pi+ electroproduction}.
\newblock {\em Eur.Phys.J.}, C65:137--151, 2010.

\bibitem{Berthou:2015oaw}
B.~Berthou, D.~Binosi, N.~Chouika, L.~Colaneri, M.~Guidal, C.~Mezrag,
  H.~Moutarde, J.~Rodríguez-Quintero, F.~Sabatié, P.~Sznajder, and J.~Wagner.
\newblock {PARTONS: PARtonic Tomography Of Nucleon Software. A computing
  framework for the phenomenology of Generalized Parton Distributions}.
\newblock {\em Eur. Phys. J.}, C78(6):478, 2018.

\bibitem{Naterer:1977fem}
F.~Natterer.
\newblock {The finite element method for ill-posed problems}.
\newblock {\em RAIRO. Analyse num\'erique}, 11(3):271--278, 1977.

\bibitem{Dutrieux:2021wll}
H.~Dutrieux, O.~Grocholski, H.~Moutarde, and P.~Sznajder.
\newblock {Artificial neural network modelling of generalised parton
  distributions}.
\newblock {\em Eur. Phys. J. C}, 82(3):252, 2022.
\newblock [Erratum: Eur.Phys.J.C 82, 389 (2022)].

\bibitem{Ilyas:2019threesigma}
I.~F. Ilyas and X.~Chu.
\newblock {\em Data Cleaning}.
\newblock Association for Computing Machinery, New York, NY, USA, 2019.

\bibitem{HornikEtAl89}
K.~Hornik, M.~Stinchcombe, and H.~White.
\newblock Multilayer feedforward networks are universal approximators.
\newblock {\em Neural Networks}, 2(5):359--366, 1989.

\bibitem{Bishop1995}
C.~M. Bishop.
\newblock Training with noise is equivalent to tikhonov regularization.
\newblock {\em Neural Comput.}, 7(1):108–116, jan 1995.

\bibitem{WagerSidaLiang2013}
S.~Wager, S.~Wang, and P.~Liang.
\newblock Dropout training as adaptive regularization.
\newblock {\em Advances in Neural Information Processing Systems}, January
  2013.
\newblock 27th Annual Conference on Neural Information Processing Systems, NIPS
  2013 ; Conference date: 05-12-2013 Through 10-12-2013.

\bibitem{DBLP:journals/corr/KingmaB14}
D.~P. Kingma and J.~Ba.
\newblock Adam: {A} method for stochastic optimization.
\newblock In Y.~Bengio and Y.~LeCun, editors, {\em 3rd International Conference
  on Learning Representations, {ICLR} 2015, San Diego, CA, USA, May 7-9, 2015,
  Conference Track Proceedings}, 2015.

\bibitem{JMLR:v15:srivastava14a}
S.~Nitish, H.~Geoffrey, K.~Alex, S.~Ilya, and S.~Ruslan.
\newblock Dropout: A simple way to prevent neural networks from overfitting.
\newblock {\em Journal of Machine Learning Research}, 15(56):1929--1958, 2014.

\bibitem{Cichy:2023dgk}
K.~Cichy et~al.
\newblock {Generalized Parton Distributions from Lattice QCD}.
\newblock {\em Acta Phys. Polon. Supp.}, 16(7):7--A6, 2023.

\bibitem{Mezrag:2016hnp}
C.~Mezrag, H.~Moutarde, and J.~Rodriguez-Quintero.
\newblock {From Bethe–Salpeter Wave functions to Generalised Parton
  Distributions}.
\newblock {\em Few Body Syst.}, 57(9):729--772, 2016.

\bibitem{Kumericki:2016ehc}
K.~Kumericki, S.~Liuti, and H.~Moutarde.
\newblock {GPD phenomenology and DVCS fitting}.
\newblock {\em Eur. Phys. J.}, A52(6):157, 2016.

\bibitem{Mezrag:2013mya}
C.~Mezrag, H.~Moutarde, and F.~Sabati\'e.
\newblock {Test of two new parameterizations of the Generalized Parton
  Distribution $H$}.
\newblock {\em Phys.Rev.}, D88:014001, 2013.

\bibitem{Mezrag:2014jka}
C.~Mezrag, L.~Chang, H.~Moutarde, C.D. Roberts, J.~Rodríguez-Quintero, et~al.
\newblock {Sketching the pion's valence-quark generalised parton distribution}.
\newblock {\em Phys.Lett.}, B741:190--196, 2014.

\bibitem{Mezrag:2014tva}
C.~Mezrag, H.~Moutarde, J.~Rodr\'iguez-Quintero, and F.~Sabati\'e.
\newblock {Towards a Pion Generalized Parton Distribution Model from
  Dyson-Schwinger Equations}.
\newblock {\em arxiv:1406.7425}, 2014.

\bibitem{Mezrag:2015mka}
C.~Mezrag.
\newblock {\em {Generalised Parton Distributions : from phenomenological
  approaches to Dyson-Schwinger equations}}.
\newblock PhD thesis, IRFU, SPhN, Saclay, 2015.

\bibitem{Chang:2013pq}
L.~Chang, I.C. Cloet, J.J. Cobos-Martinez, C.D. Roberts, S.M. Schmidt, et~al.
\newblock {Imaging dynamical chiral symmetry breaking: pion wave function on
  the light front}.
\newblock {\em Phys.Rev.Lett.}, 110:132001, 2013.

\bibitem{Nakanishi:1963zz}
N.~Nakanishi.
\newblock {Partial-Wave Bethe-Salpeter Equation}.
\newblock {\em Phys.Rev.}, 130:1230--1235, 1963.

\bibitem{Nakanishi:1969ph}
N.~Nakanishi.
\newblock {A General survey of the theory of the Bethe-Salpeter equation}.
\newblock {\em Prog.Theor.Phys.Suppl.}, 43:1--81, 1969.

\bibitem{Kroll:2012sm}
P.~Kroll, H.~Moutarde, and F.~Sabatie.
\newblock {From hard exclusive meson electroproduction to deeply virtual
  Compton scattering}.
\newblock {\em Eur.Phys.J.}, C73:2278, 2013.

\bibitem{Musatov:1999xp}
I.V. Musatov and A.V. Radyushkin.
\newblock {Evolution and models for skewed parton distributions}.
\newblock {\em Phys.Rev.}, D61:074027, 2000.

\bibitem{Pumplin:2002vw}
J.~Pumplin, D.R. Stump, J.~Huston, H.L. Lai, Pavel~M. Nadolsky, et~al.
\newblock {New generation of parton distributions with uncertainties from
  global QCD analysis}.
\newblock {\em JHEP}, 0207:012, 2002.

\end{thebibliography}

\end{document}